\documentclass[prd,amsmath,amssymb, aps,superscriptaddress, nofootinbib]{revtex4}
\usepackage[version=4]{mhchem} 
\usepackage[normalem]{ulem}
\usepackage{amsfonts}
\allowdisplaybreaks[4]
\usepackage{mathrsfs}
\usepackage{graphicx}
\usepackage{xcolor}
\usepackage{xfrac}
\usepackage{footmisc}
\usepackage{pifont}
\usepackage{times}
\usepackage{epstopdf}
\usepackage{enumerate}
\usepackage[hidelinks]{hyperref}
\usepackage{enumitem}
\usepackage{multirow}
\usepackage{booktabs}
\usepackage{slashed}
\usepackage{relsize}
\usepackage{epsfig}
\usepackage{verbatim}
\usepackage{bm}
\usepackage[utf8]{inputenc}
\usepackage{graphics}
\usepackage{subfigure}
\usepackage[english]{babel}
\usepackage{orcidlink}

\definecolor{zima_blue}{HTML}{1393C1}
\hypersetup{setpagesize=false,
bookmarksnumbered=true,
bookmarksopen=true, 
colorlinks=true,
linkcolor=red,
urlcolor=blue,
citecolor=blue,
linktocpage=false}
\begin{document}
\newcommand{\psl}{ p \hspace{-1.8truemm}/ }
\newcommand{\nsl}{ n \hspace{-2.2truemm}/ }
\newcommand{\vsl}{ v \hspace{-2.2truemm}/ }
\newcommand{\epsl}{\epsilon \hspace{-1.8truemm}/\,  }

\title{ Revisiting $\bar B^0 \rightarrow \Lambda_c^+ \bar p$ decay with higher twist corrections }

\author{Zhi-Tian Zou} 
\email{zouzt@ytu.edu.cn}
\affiliation{Department of Physics, Yantai University, Yantai 264005, China}

\author{Ying Li} 
\email{liying@ytu.edu.cn}
\affiliation{Department of Physics, Yantai University, Yantai 264005, China}

\author{Zhou Rui} 
\email{jindui1127@126.com}
\affiliation{Department of Physics, Yantai University, Yantai 264005, China}

\date{\today}
\begin{abstract}
We investigate the single-charmed baryonic decays $\bar B^0 \to \Lambda_c^+ \bar p$ and $\bar B^0 \to \bar\Lambda_c^- p$, which receive contributions from both $W$-emission and $W$-exchange topologies, within the framework of perturbative QCD. Higher-power corrections associated with the hadronic light-cone distribution amplitudes of both the initial- and final-state hadrons are systematically taken into account. We find that these higher-twist contributions play an important role in baryonic $B$ decays and cannot be neglected. A sizable destructive interference between the $W$-emission and $W$-exchange amplitudes is observed, which significantly reduces the predicted branching fraction of $\bar B^0 \to \Lambda_c^+ \bar p$ and leads to improved agreement with experimental measurements. The doubly Cabibbo-suppressed decay $\bar B^0 \to \bar\Lambda_c^- p$ is studied for the first time. 
Its branching fraction is predicted to be of order $10^{-8}$, placing it within the reach of future high-luminosity experiments. We further present the first theoretical predictions for the decay asymmetry parameters of both channels, which provide additional observables for testing the underlying decay dynamics and can be confronted with future experimental data.
\end{abstract}
\maketitle

\section{Introduction}
The QCD dynamics of baryonic $B$ decays are considerably more intricate than those of mesonic decays and remain only partially understood from first principles. Experimentally, two-body baryonic $B$ decays are generally suppressed relative to multibody channels, a feature commonly attributed to the threshold enhancement effect~\cite{Hou:2000bz}. This suggests that the $B$ meson preferentially hadronizes into a baryon--antibaryon pair with a relatively small energy release. In particular, charmful baryonic $B$ decays induced by the $b\to c$ transition exhibit a rich and nontrivial pattern, providing a useful laboratory for studying the interplay between weak decay mechanisms and nonperturbative QCD dynamics.

A representative example is the decay $\bar{B}\to\bar{\Lambda}_c\Xi_c$, which proceeds via the $b\to c\bar{c}s$ transition. Its branching fraction is measured to be of order $10^{-3}$~\cite{Belle:2018kzz,Belle:2019bgi}, and is well reproduced by various theoretical approaches~\cite{Cheng:2005vd,Hsiao:2023mud,Cheng:2009yz,Rui:2024xgc,Geng:2025yna,Geng:2024uxp,Chua:2026awd}. In contrast, the related decay $\bar B^0\to \Lambda_c^+\bar p$, governed by the $b\to c\bar{u}d$ transition, has a branching fraction at the level of $10^{-5}$~\cite{Belle:2002gir,BaBar:2008get}. Since the Cabibbo--Kobayashi--Maskawa (CKM) factors involved in these two processes are of comparable magnitude, the observed two-order-of-magnitude difference in their decay rates must originate from nontrivial QCD dynamics.

This hierarchy can be qualitatively understood through the $Q$-value mechanism, where $Q$ denotes the mass difference between the initial and final states~\cite{BaBar:2008get}. Empirically, baryonic $B$ decay rates tend to increase as the available phase space decreases. Because the $Q$ value is smallest for decays into two charmed baryons and significantly larger when a charmless baryon is present, $\bar B^0\to \Lambda_c^+\bar p$ is expected to be suppressed relative to $\bar B\to\bar{\Lambda}_c\Xi_c$. A more dynamical interpretation further indicates that different topological requirements for hard gluon exchanges lead to distinct suppression patterns~\cite{Cheng:2005vd}. In particular, while the production of an energetic $\bar{\Lambda}_c\Xi_c$ pair does not require additional hard gluon exchange, the $\Lambda_c^+\bar p$ channel typically involves two extra hard gluons, resulting in a suppression factor of order $\alpha_s^4\sim10^{-2}$~\cite{Cheng:2005vd}. Despite its qualitative success, a quantitative description of absolute branching fractions remains challenging, and systematic QCD-based analyses are still limited.

The decay $\bar B^0\to \Lambda_c^+\bar p$ has been studied extensively using a variety of phenomenological approaches, including the pole model~\cite{Jarfi:1990ej,Cheng:2002sa,Cheng:2001ub}, diquark models~\cite{Ball:1990fw}, QCD sum rules~\cite{Chernyak:1990ag}, and flavor-symmetry-based methods~\cite{Li:1989vp,Savage:1989jx,He:1989re,Luo:2003pv}. Early estimates typically predicted branching fractions of order $10^{-3}$, which significantly exceed experimental measurements~\cite{ParticleDataGroup:2026aaa}. More recent studies, including updated pole-model analyses~\cite{Cheng:2002sa,Cheng:2001ub} and alternative treatments~\cite{Hsiao:2019wyd}, obtain values at the level of $10^{-5}$, which are closer to data but still incomplete in their dynamical content. Within the perturbative QCD (PQCD) framework, a previous calculation including only the $W$-emission topology yields $(2.3\text{--}5.1)\times10^{-5}$~\cite{He:2006vz}, deviating from the experimental central value by more than four standard deviations. Given that this is the most precisely measured two-body baryonic $B$ decay, a fully consistent theoretical description remains an open problem.

Motivated by these issues, we perform a comprehensive PQCD analysis of $\bar B^0\to \Lambda_c^+\bar p$. The decay receives contributions from both $W$-emission and $W$-exchange topologies. In our previous study of doubly charmed baryonic $B$ decays~\cite{Rui:2024xgc}, we found that $W$-exchange contributions are not strongly helicity suppressed due to the presence of the heavy charm quark, suggesting that they may also play an important role in single-charmed baryonic modes. In the present work, we therefore include all relevant topological amplitudes, going beyond earlier PQCD analyses that considered only $W$-emission contributions.

We further incorporate higher-twist effects arising from the light-cone distribution amplitudes (LCDAs) of both initial- and final-state hadrons. To assess theoretical uncertainties associated with baryonic structure, we construct three alternative models for the $\Lambda_c$ LCDAs based on heavy-quark symmetry. In addition, we extend our analysis to the doubly Cabibbo-suppressed decay $\bar B^0\to \bar\Lambda_c^- p$, which has not been previously studied  and may be accessible in future high-statistics experiments. These investigations aim to provide a more coherent and systematically improved description of two-body baryonic $B$ decays within a QCD-based framework.

The remainder of this paper is organized as follows. In Sec.~\ref{sec:framework}, we present the effective Hamiltonian, kinematic setup, and hadronic LCDAs. Section~\ref{sec:results} contains the analysis of the decay amplitudes, including all higher-power contributions, together with numerical results for branching fractions and asymmetry parameters. A summary is given in Sec.~\ref{sec:sum}, and technical details of the factorization formulas are collected in Appendix~\ref{sec:for}.

\section{Theoretical framework}\label{sec:framework}

\begin{figure*}[htbp]
\centering
\includegraphics[width=0.45\textwidth]{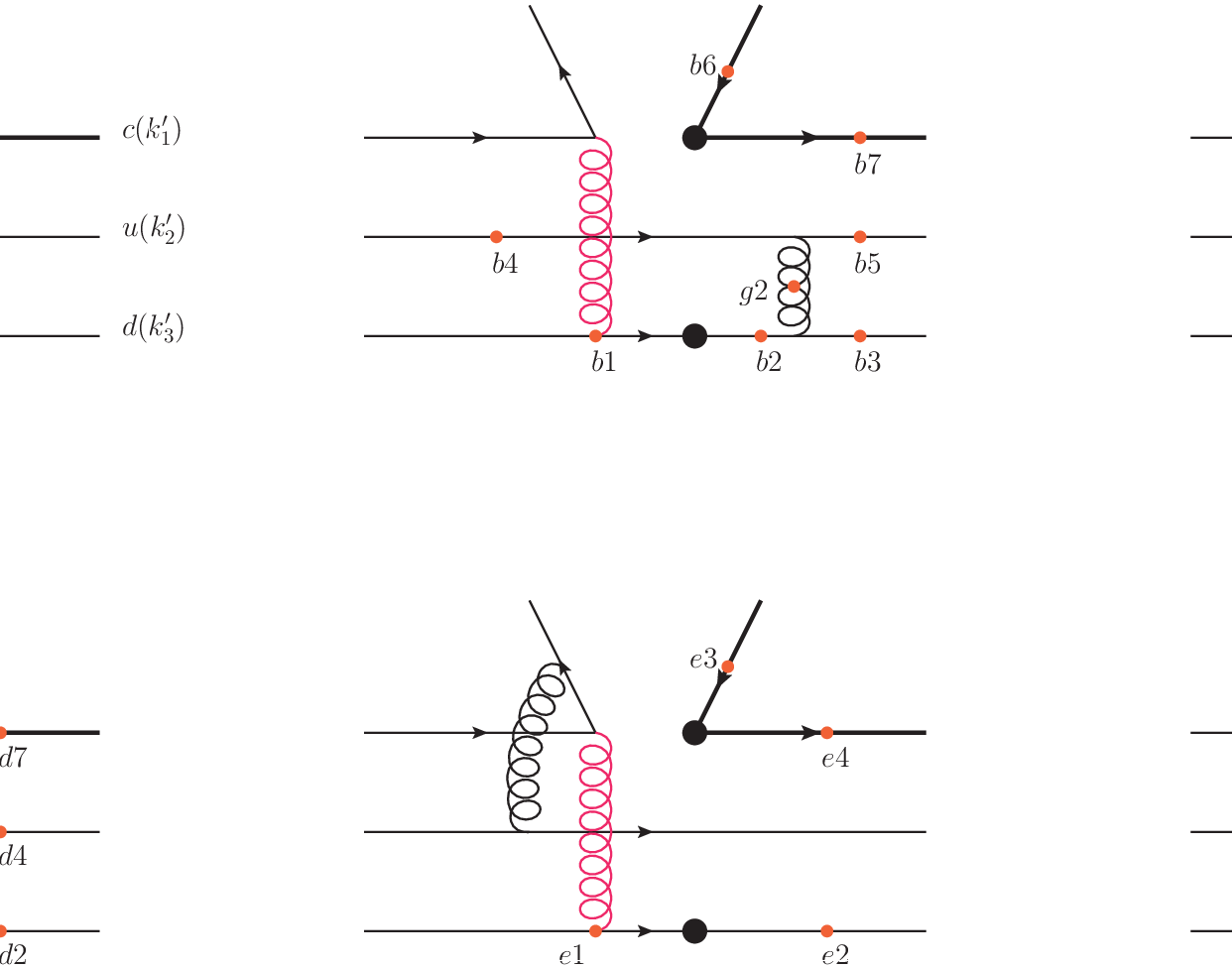}
\caption{$W$-emission diagrams for the $\bar B^0\rightarrow \Lambda_c^+ \bar p$  decay at leading order. The solid black blobs represent the vertex of the effective weak interaction, while the red ones indicate the possible connections of the gluon (red) attached to the spectator $\bar d$ quark.
 The heavy quarks $b$ and $c$ are shown in bold lines.}
\label{fig:C}
\end{figure*}

\begin{figure*}[htbp]
\centering
\includegraphics[width=0.45\textwidth]{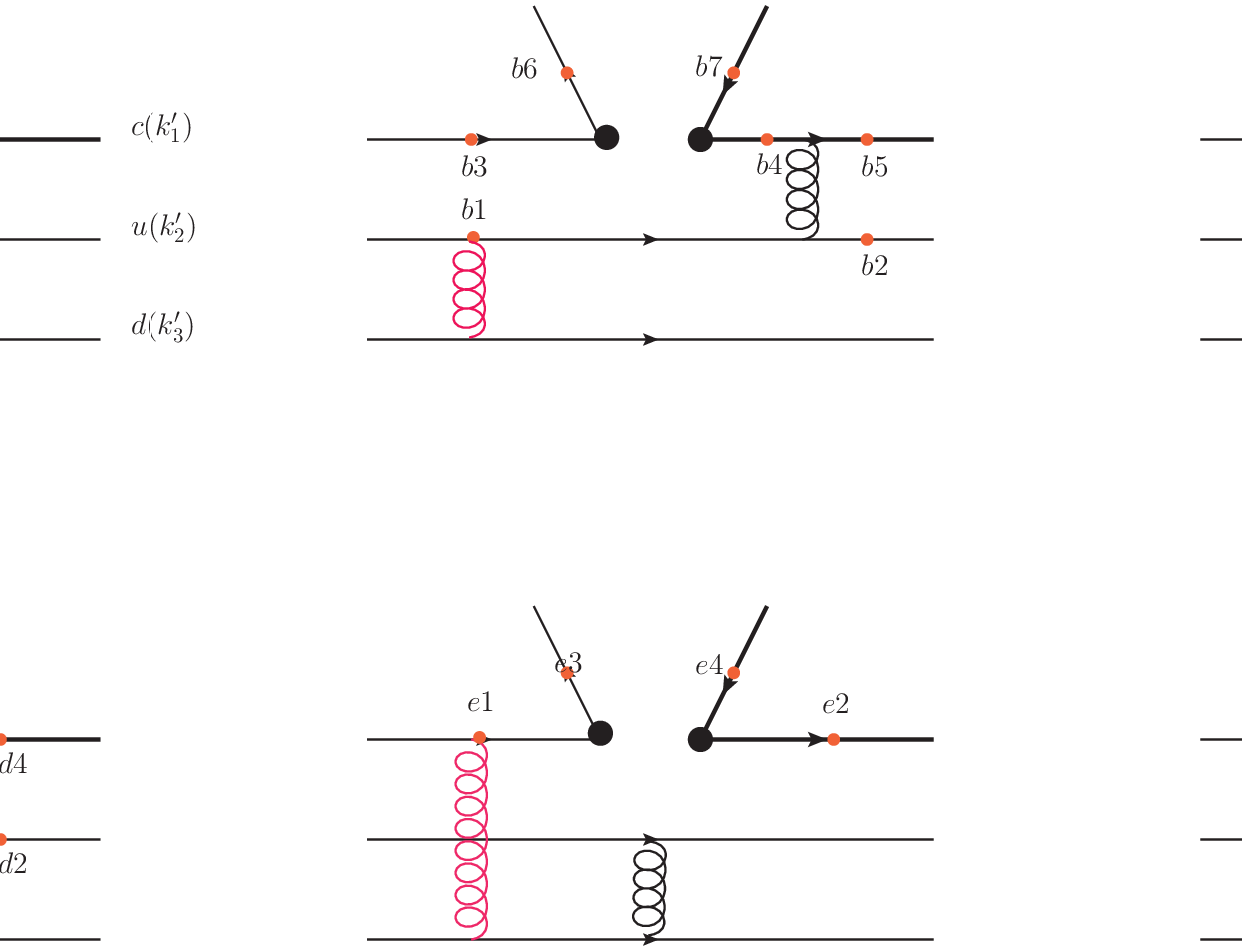}
\caption{$W$-exchange diagrams for the $\bar B^0\rightarrow \Lambda_c^+ \bar p$  decay at leading order.}
\label{fig:E}
\end{figure*}

The decay $\bar B^0\to \Lambda_c^+\bar p$ receives contributions from both $W$-emission and $W$-exchange topologies, as shown in Figs.~\ref{fig:C} and~\ref{fig:E}, respectively. The corresponding Feynman diagrams for $\bar B^0\to \bar \Lambda_c^- p$ are obtained by interchanging the $c$ and $u$ quarks at the effective weak vertex. The calculation is performed in the rest frame of the $B$ meson. In light-cone coordinates, the $B$-meson momentum is written as $q = \frac{M}{\sqrt{2}}(1,1,\mathbf{0}_T)$, where $M$ denotes the $B$-meson mass. Throughout this work, the antibaryon (baryon) is assumed to move along the plus (minus) light-cone direction with momentum $p^{(\prime)}$, given by
\begin{eqnarray}\label{eq:pq}
p=\frac{M}{\sqrt{2}}(f^+,f^-,\mathbf{0}_{T}),\quad
p'=\frac{M}{\sqrt{2}}(1-f^+,1-f^-,\mathbf{0}_{T}).
\end{eqnarray}
The variables $f^\pm$ are defined as
\begin{eqnarray}
f^\pm=\frac{1}{2}\left(1-r^2+\bar r^2 \pm \sqrt{(1-r^2+\bar r^2)^2-4\bar r^2}\right),
\end{eqnarray}
where the mass ratios are $r(\bar r)=m(\bar m)/M$, with $m$ ($\bar m$) denoting the mass of the baryon (antibaryon).

\vspace{10mm}

To evaluate the hard kernels for $\bar B^0\to \Lambda_c^+\bar p$, we parametrize the valence-quark momenta in the first diagram of Fig.~\ref{fig:C} as\footnote{For $\bar B^0\to \Lambda_c^+\bar p$, $p'$ and $p$ correspond to the momenta of $\Lambda_c^+$ and $\bar p$, respectively.}
\begin{align}\label{eq:dyn}
q_1=\left(0,\frac{M}{\sqrt{2}}y,\mathbf{q}_{T}\right),&\quad q_2=q-q_1,\quad
k_i=\left(\frac{M}{\sqrt{2}}f^+x_i,0,\mathbf{k}_{iT}\right), \nonumber\\
k_1'=p'-k_2'-k_3',\quad
k_2'&=\left(0,\frac{M}{\sqrt{2}}(1-f^-)x_2',\mathbf{k}'_{2T}\right),\quad
k_3'=\left(0,\frac{M}{\sqrt{2}}(1-f^-)x_3',\mathbf{k}'_{3T}\right),
\end{align}
where $y$, $x_l$, and $x_l'$ ($l=1,2,3$) denote the longitudinal momentum fractions carried by the constituent quarks, and $\mathbf{q}_T$, $\mathbf{k}_{lT}$, and $\mathbf{k}'_{lT}$ are the corresponding transverse momenta. The $b$ and $c$ quarks are treated as massive and carry momenta $q_2$ and $k_1'$, respectively, while the masses of light quarks are neglected. For the decay $\bar B^0\to \bar \Lambda_c^- p$, the two final-state baryons are interchanged, with $p$ and $p'$ now denoting the momenta of $\bar \Lambda_c^-$ and $p$, respectively. Maintaining the definitions of $p$ and $p'$ in Eq.~(\ref{eq:pq}), the valence-quark momenta are parametrized as follows. For the internal $W$-emission diagrams, we set
\begin{align}\label{eq:dyn1}
k_l'=\left(0,\frac{M}{\sqrt{2}}(1-f^-)x_l',\mathbf{k}'_{lT}\right),&\quad
k_1=\left(\frac{M}{\sqrt{2}}f^+x_1,0,\mathbf{k}_{1T}\right),\nonumber\\
k_2=\left(\frac{M}{\sqrt{2}}f^+x_2,0,\mathbf{k}_{2T}\right),&\quad
k_3=p-k_1-k_2,
\end{align}
while for the $W$-exchange diagrams, we have
\begin{align}\label{eq:dyn2}
k_l'=\left(0,\frac{M}{\sqrt{2}}(1-f^-)x_l',\mathbf{k}'_{lT}\right),&\quad
k_1=p-k_2-k_3,\nonumber\\
k_2=\left(\frac{M}{\sqrt{2}}f^+x_2,0,\mathbf{k}_{2T}\right),&\quad
k_3=\left(\frac{M}{\sqrt{2}}f^+x_3,0,\mathbf{k}_{3T}\right).
\end{align}

In the calculation of the decay amplitudes, hadronic wave functions are defined through nonlocal matrix elements of quark bilinear operators sandwiched between hadronic states and the vacuum. We now introduce the wave functions relevant to this work. Following Refs.~\cite{Grozin:1996pq,Beneke:2000ry}, the $B$-meson wave function is defined as
\begin{eqnarray}\label{eq:bb}
\Phi_{B} &=& \frac{i}{\sqrt{2N_c}}(\slashed{q}+M)\gamma_5\left(\frac{\slashed{n}_+}{\sqrt{2}}\phi^+_{B} + \frac{\slashed{n}_-}{\sqrt{2}}\phi^-_{B}\right)
\nonumber\\
&=& -\frac{i}{\sqrt{2N_c}}(\slashed{q}+M)\gamma_5\left(\phi_{B} + \frac{\slashed{n}_+}{\sqrt{2}} \bar{\phi}_B\right),
\end{eqnarray}
where $\phi_{B}=\phi^-_{B}$ and $\bar{\phi}_B=\phi^-_{B}-\phi^+_{B}$. These light cone distribution amplitudes encode the nonperturbative bound-state dynamics of the $B$ meson. The lightlike vectors $n_+=(1,0,\mathbf{0}_T)$ and $n_-=(0,1,\mathbf{0}_T)$ specify the plus and minus directions, respectively. The two distribution amplitudes $\phi_B^{\pm}$ have distinct asymptotic behaviors but satisfy the same normalization condition,
\begin{eqnarray}\label{eq:nou}
\int_0^1 \phi^{\pm}_{B}(y) dy = \frac{f_{B}}{2\sqrt{2N_c}}.
\end{eqnarray}
In this work, we adopt   a conventional ansatz  for $\phi_B^{-}$,
\begin{eqnarray}\label{eq:gaus}
\phi^-_{B}(y,b_q)=N_B y^2(1-y)^2 \exp\left[-\frac{y^2M^2}{2\omega_b^2}-\frac{\omega_b^2 b_q^2}{2}\right],
\end{eqnarray}
where $N_B$ is the normalization constant determined by Eq.~(\ref{eq:nou}), and $b_q$ denotes the variable conjugate to the parton transverse momentum $\mathbf{q}_T$ in the $B$ meson. The shape parameter is taken to be $\omega_b = 0.4~\mathrm{GeV}$ for the $B^0$ meson~\cite{Hua:2020usv}.  The parametrization in Eq.~(\ref{eq:gaus}) combines the QCD asymptotic behavior $\phi_B(y) \sim y^2(1-y)^2$ near the endpoints with a Gaussian factor that models the transverse momentum dependence and provides a smooth suppression at large $y$. This form has been extensively used in PQCD analyses of $B$ decays~\cite{Hua:2020usv,Kurimoto:2006iv, Yang:2020xal}  and provides a reasonable phenomenological description of the nonperturbative $B$-meson structure.
The distribution amplitude $\phi_B^{+}$ can be obtained from the equation of motion relation (neglecting three-parton contributions)~\cite{Grozin:1996pq},
\begin{eqnarray}
\phi^-_{B} = -y \frac{d\phi^+_{B}}{dy},
\end{eqnarray}
and its explicit form can be found in Refs.~\cite{Kurimoto:2006iv, Yang:2020xal}.

The LCDAs of charmed baryons have been studied less extensively in the literature. In this work, we assume that, in the heavy-quark limit, their structures can be approximated by those of bottom baryons. For a detailed discussion of bottom-baryon LCDAs, we refer to Refs.~\cite{Ball:2008fw, Bell:2013tfa, Ali:2012pn, Braun:2014npa, Wang:2015ndk, Ali:2012zza}. The nonlocal matrix element defining the $\Lambda_c$ LCDA is written as~\cite{Rui:2024xgc}
\begin{eqnarray}
\epsilon^{ijk}\langle \Lambda_c |\bar {u}_{\alpha}^i(t_1)\bar {d}_{\beta}^j(t_2)\bar {c}_{\gamma}^k(0)|0\rangle&=&
\frac{ f^{(1)}_{\Lambda_c}}{8} \bar u_{\gamma}
[(C\gamma_5\slashed{\bar n})_{\beta\alpha}\Phi_{2}(t_1,t_2)+(C\gamma_5\slashed{ n})_{\beta\alpha}\Phi_4(t_1,t_2)] \nonumber\\
&&+\frac{f^{(2)}_{\Lambda_c}}{4}\bar u_{\gamma}[(C\gamma_5)_{\beta\alpha}\Phi_3^s(t_1,t_2)+\frac{i}{2}(C\gamma_5\sigma_{\bar nn})_{\beta\alpha}\Phi_3^a(t_1,t_2)].
\end{eqnarray}
where $\epsilon^{ijk}$ is the totally antisymmetric tensor in color space, and $i,j,k$ are color indices. The indices $\alpha,\beta,\gamma$ denote Dirac components, and $C$ is the charge-conjugation matrix. The light cone vectors $n$ and $\bar n$ satisfy $n^2=\bar n^2=0$ and $n\cdot\bar n=2$. The field $c(0)$ denotes the static charm-quark field at the origin, while $u(t_1)$ and $d(t_2)$ are light-quark fields separated from the origin along the light-like directions. The spinor $u$ denotes the $\Lambda_c$ baryon spinor satisfying $\slashed{p}'u = m_{\Lambda_c} u$, where $m_{\Lambda_c}$ is the $\Lambda_c$ mass. For the decay constants, we adopt $f^{(1)}_{\Lambda_c}=f^{(2)}_{\Lambda_c}=0.022\pm0.008~\text{GeV}^3$ from QCD sum-rule estimates~\cite{Wang:2010fq}. The functions $\Phi_{2,3,4}$ denote the $\Lambda_c$ LCDAs of twist 2, 3, and 4, respectively. SU(3) symmetry implies that $\Phi_3^a$ is antisymmetric under the interchange of the $u$ and $d$ quarks, whereas $\Phi_2$, $\Phi_4$, and $\Phi_3^s$ are symmetric under the same operation. Their normalization conditions are given by $\Phi_3^a(0,0)=0$ and $\Phi_2(0,0)=\Phi_4(0,0)=\Phi_3^s(0,0)=1$. Transforming to momentum space, the LCDAs are defined as~\cite{Ball:2008fw}
\begin{eqnarray}
\Phi(t_1,t_2)=\int_0^\infty d\omega\omega \int_0^1 du \phi(\omega,u)e^{-i\omega[t_1 u+t_2(1-u)]},
\end{eqnarray}
where $\omega$ denotes the total energy carried by the two light quarks in the heavy-quark rest frame, and $u$ is the momentum fraction carried by the $u$ quark inside the light-diquark system. These variables are related to partonic momentum fractions by $\omega=(x_2+x_3)m_{\Lambda_c}$ and $u=x_2/(x_2+x_3)$, where $x_2$ and $x_3$ are the longitudinal momentum fractions of the $u$ and $d$ quarks inside the $\Lambda_c$ baryon, respectively.

Following the approach in Ref.~\cite{Rui:2025iwa}, we construct three representative models for $\phi(\omega,u)$:
\begin{itemize}
\item Exponential model~\cite{Bell:2013tfa}:
\begin{eqnarray}
\phi_2(\omega,u)&=&   N_2 \mathcal{T}(x_1,x_2,x_3) \frac{\omega^2u(1-u)}{\omega_0^4}e^{-\frac{\omega}{\omega_0}},\nonumber\\
\phi_3^{s}(\omega,u)&=&N_3 \mathcal{T}(x_1,x_2,x_3)\frac{\omega}  {2\omega_0^3}e^{-\frac{\omega}{\omega_0}},\nonumber\\
\phi_3^{a}(\omega,u)&=&N_3 \mathcal{T}(x_1,x_2,x_3)\frac{\omega(2u-1)}  {2\omega_0^3}e^{-\frac{\omega}{\omega_0}},\nonumber\\
\phi_4(\omega,u)&=&    N_4 \mathcal{T}(x_1,x_2,x_3) \frac{1}     {\omega_0^2}e^{-\frac{\omega}{\omega_0}},
\end{eqnarray}
where $\omega_0=0.4\pm0.1$ GeV characterizes the average energy carried by the two light quarks inside the $\Lambda_c$ baryon. The common factor $\mathcal{T}(x_1,x_2,x_3)= e^{-\frac{m_{\Lambda_c}^2}{2\beta^2x_1}-\frac{m_u^2}{2\beta^2x_2}-\frac{m_d^2}{2\beta^2x_3}}$, with $m_u=m_d=0.22$ GeV~\cite{Li:2021qod} and $\beta=1.0$ GeV~\cite{Shih:1998pb}, is introduced to ensure that the heavy charm quark remains close to its mass shell~\cite{Rui:2025iwa}.

\item QCD sum-rule (QCDSR) model~\cite{Ball:2008fw}:
\begin{eqnarray}
\phi_2    (\omega,u)&=&    N_2 \mathcal{T}(x_1,x_2,x_3) \frac{15}{2\mathcal{N}}\omega^2 u (1-u) \int_{\frac{\omega}{2}}^{s_0}dse^{-s/\tau}(s-\frac{\omega}{2}),\nonumber\\
\phi_3^{s}(\omega,u)&=&    N_3 \mathcal{T}(x_1,x_2,x_3) \frac{15}{4\mathcal{N}}\omega           \int_{\frac{\omega}{2}}^{s_0}dse^{-s/\tau}(s-\frac{\omega}{2})^2,\nonumber\\
\phi_3^{a}(\omega,u)&=&    N_3 \mathcal{T}(x_1,x_2,x_3) \frac{15}{4\mathcal{N}}\omega  (2u-1)   \int_{\frac{\omega}{2}}^{s_0}dse^{-s/\tau}(s-\frac{\omega}{2})^2,\nonumber\\
\phi_4    (\omega,u)&=&    N_4 \mathcal{T}(x_1,x_2,x_3) \frac{5}{\mathcal{N}}                   \int_{\frac{\omega}{2}}^{s_0}dse^{-s/\tau}(s-\frac{\omega}{2})^3,
\end{eqnarray}
where $\mathcal{N}=\int_0^{s_0}dse^{-s/\tau}s^5$, with continuum threshold $s_0=1.2$ GeV and Borel parameter $\tau\in [0.4,0.8]$ GeV. Note that the variable $\omega$ is restricted to the interval $0<\omega<2s_0$ when performing the integrations~\cite{Ball:2008fw}.

\item Gegenbauer model~\cite{Ali:2012pn}:
\begin{eqnarray}\label{eq:twust}
\phi_2  (\omega,u)&=&  N_2 \mathcal{T}(x_1,x_2,x_3) \omega^2u(1-u) \sum_{l=0}^2\frac{a_l}{\epsilon_l^4}\frac{c_l^{3/2}(2u-1)}{|c_l^{3/2}|^2}e^{-\frac{\omega}{\epsilon_l}},\nonumber\\
\phi_3^{s,a}(\omega,u)&=& N_3 \mathcal{T}(x_1,x_2,x_3) \frac{\omega}{2}  \sum_{l=0}^2\frac{a_l}{\epsilon_l^3}\frac{c_l^{1/2}(2u-1)}{|c_l^{1/2}|^2}e^{-\frac{\omega}{\epsilon_l}},\nonumber\\
\phi_4  (\omega,u)&=&  N_4 \mathcal{T}(x_1,x_2,x_3) \sum_{l=0}^2\frac{a_l}{\epsilon_l^2}\frac{c_l^{1/2}(2u-1)}{|c_l^{1/2}|^2}e^{-\frac{\omega}{\epsilon_l}},
\end{eqnarray}
with the Gegenbauer polynomials defined as
\begin{eqnarray}
C_0^{\xi}(x)&=&1, \quad C_1^{\xi}(x)=2\xi x, \quad C_2^{\xi}(x)=2\xi(1+\xi)x^2-\xi.
\end{eqnarray}
The shape parameters $a_l$ and $\epsilon_l$ depend on a free parameter $A=0.5\pm 0.2$, and their explicit expressions can be found in Ref.~\cite{Ali:2012pn}.
\end{itemize}

It should be noted that the normalization constants $N_i$ differ for each model and are fixed by the normalization condition
\begin{eqnarray}
\int^\infty_0 d\omega \int^1_0 du\omega \phi_{i}(\omega,u) =1.
\end{eqnarray}
The three models for the $\Lambda_c$ LCDAs represent different levels of theoretical input. The Exponential model~\cite{Bell:2013tfa} is a simple phenomenological parametrization based on HQET. The QCDSR model~\cite{Ball:2008fw} is derived from QCD sum rules and incorporates continuum threshold and Borel parameters, providing a more QCD-motivated shape. The Gegenbauer model~\cite{Ali:2012pn} expands the LCDAs in conformal partial waves, allowing for systematic inclusion of higher moments.  In the numerical section~\ref{sec:results}, we will provide a detailed comparison of the effects of these three models on the physical observables under consideration.  

The proton  LCDAs can also be defined through matrix elements of nonlocal three-quark operators sandwiched between the vacuum and the hadronic state~\cite{Braun:2000kw}. For an outgoing antiproton state, one has
 \begin{eqnarray}\label{eq:LCDAsp}
4\langle \bar p|\varepsilon^{ijk}u_\alpha^i(a_1z)u_\beta^j(a_2z)d_\gamma^k(a_3z)|0\rangle &=& \mathcal{S}_1m_pC_{\beta\alpha}(\gamma_5v)_\gamma-
\mathcal{S}_2m^2_pC_{\beta\alpha}(\slashed{z}\gamma_5v)_\gamma+  \mathcal{P}_1m_p(\gamma_5C)_{\beta\alpha}(v)_\gamma\nonumber\\&&
- \mathcal{P}_2m^2_p(\gamma_5C)_{\beta\alpha}(\slashed{z}v)_\gamma +\mathcal{V}_1(\slashed{p}C)_{\beta\alpha}(\gamma_5v)_\gamma
-\mathcal{V}_2m_p(\slashed{p}C)_{\beta\alpha}(\slashed{z}\gamma_5v)_\gamma \nonumber\\&& -\mathcal{V}_3m_p(\gamma_\mu C)_{\beta\alpha}( \gamma^\mu\gamma_5v)_\gamma
+\mathcal{V}_4m^2_p(\slashed{z} C)_{\beta\alpha}(\gamma_5v)_\gamma +\mathcal{V}_5m^2_p(\gamma_\mu C)_{\beta\alpha}(i \sigma^{\mu\nu}z_{\nu}\gamma_5v)_\gamma \nonumber\\&&
-\mathcal{V}_6m^3_p(\slashed{z}C)_{\beta\alpha}(\slashed{z}\gamma_5v)_\gamma +\mathcal{A}_1(\gamma_5\slashed{p}C)_{\beta\alpha}(N)_\gamma
-\mathcal{A}_2m_p(\gamma_5\slashed{p}C)_{\beta\alpha}(\slashed{z}v)_\gamma \nonumber\\&&-\mathcal{A}_3m_p(\gamma_5\gamma_\mu C)_{\beta\alpha}(\gamma^\mu v)_\gamma
+\mathcal{A}_4m^2_p(\gamma_5\slashed{z}C)_{\beta\alpha}( v)_\gamma +\mathcal{A}_5m^2_p(\gamma_5\gamma_\mu C)_{\beta\alpha}(i \sigma^{\mu\nu}z_{\nu} v)_\gamma \nonumber\\&&
-\mathcal{A}_6m^3_p(\gamma_5\slashed{z}C)_{\beta\alpha}(\slashed{z} v)_\gamma +\mathcal{T}_1(i p^\nu \sigma_{\mu\nu}C)_{\beta\alpha}(  \gamma^{\mu} \gamma_5 v)_\gamma
+\mathcal{T}_2m_p(i z^\mu p^\nu \sigma_{\mu\nu}C)_{\beta\alpha}( \gamma_5 v)_\gamma \nonumber\\&&
-\mathcal{T}_3m_p( \sigma_{\mu\nu}C)_{\beta\alpha}(\sigma^{\mu\nu} \gamma_5 v)_\gamma -\mathcal{T}_4m_p( p^\nu \sigma_{\mu\nu}C)_{\beta\alpha}(\sigma^{\mu\rho}z_\rho \gamma_5 v)_\gamma \nonumber\\&&
+\mathcal{T}_5m^2_p( iz^\nu\sigma_{\mu\nu}C)_{\beta\alpha}(\gamma^{\mu} \gamma_5 v)_\gamma +\mathcal{T}_6m^2_p( iz^\mu p^\nu \sigma_{\mu\nu}C)_{\beta\alpha}(\slashed{z} \gamma_5 v)_\gamma \nonumber\\&&
+\mathcal{T}_7m^2_p( \sigma_{\mu\nu}C)_{\beta\alpha}(\sigma^{\mu\nu}\slashed{z} \gamma_5 v)_\gamma -\mathcal{T}_8m^3_p( z^\mu \sigma_{\mu\nu}C)_{\beta\alpha}(  \sigma^{\mu\rho}z_\rho\gamma_5 v)_\gamma,
\end{eqnarray}
where $v$ denotes the antiproton spinor satisfying the Dirac equation $\slashed{p}v=-m_p v$, with $p$ and $m_p$ being the proton momentum and mass, respectively. The parameters $a_i$ are real numbers specifying the light-cone positions of the valence quarks, and $z$ is an arbitrary light-like vector with $z^2=0$. In this work, we choose $z = \frac{\sqrt{2}}{M f^+} n_-$ such that $z\cdot p=1$~\cite{Braun:2000kw}. It should be noted that the calligraphic invariant functions $\mathcal{F}_i$ ($\mathcal{F}=\mathcal{S},\mathcal{P},\mathcal{V},\mathcal{A},\mathcal{T}$) in Eq.~(\ref{eq:LCDAsp}) do not correspond to definite twist. They can be systematically decomposed into 24 proton LCDAs with well-defined twist and symmetry properties, which depend on the longitudinal momentum fractions carried by the constituent quarks. The explicit relations and expressions of these LCDAs can be found in Ref.~\cite{Braun:2000kw} and are not repeated here.    A conformal expansion of these LCDAs introduces 21 nonperturbative parameters, which can be further reduced, by employing isospin symmetry and the equations of motion, to    eight independent ones,  defined as~\cite{Braun:2000kw}:
\begin{align}\label{eq:fn}
\langle 0 | \varepsilon^{ijk} \left[ u^i(0) C \slashed{z} u^j(0) \right] \gamma_5 \slashed{z} d^k(0) | p \rangle &= f_{N} p z \slashed{z} N(p), \nonumber\\
\langle 0 | \varepsilon^{ijk} \left[ u^i(0) C \gamma_\mu u^j(0) \right] \gamma_5 \gamma^\mu d^k(0) | p \rangle &= \lambda_1 m_p N(p), \nonumber\\
\langle 0 | \varepsilon^{ijk} \left[ u^i(0) C \sigma_{\mu\nu} u^j(0) \right] \gamma_5 \sigma^{\mu\nu} d^k(0) | p \rangle &= \lambda_2 m_p N(p),\nonumber\\
\langle 0 | (u(0)C \slashed{z} u(0))  \gamma_5 \slashed{z} (iz \overrightarrow{D} d)(0)  |p\rangle &= f_N V_1^d (pz)^2  \slashed{z} N(p),\nonumber\\
\langle 0 | \left( u(0)C \slashed{z} \gamma_5 iz \overleftrightarrow{D} u(0) \right) \slashed{z} d(0)  |p\rangle &= -f_N A_1^u (pz)^2  \slashed{z} N(p), \nonumber\\
\langle 0 | (u(0)C \gamma_\mu u(0))  \gamma_5 \gamma^\mu \slashed{z} (iz \overrightarrow{D} d)(0) |p\rangle &= \lambda_1 f_1^d (pz) m_p  \slashed{z} N(p), \nonumber\\
\langle 0 | (u^a(0)C \sigma_{\mu\nu} u(0))  \gamma_5 \sigma^{\mu\nu} \slashed{z} (iz \overrightarrow{D} d)(0)  |p\rangle &= -\lambda_2 f_2^d (pz) m_p \slashed{z} N(p), \nonumber\\
\langle 0 | \left( u(0)C \gamma_\mu \gamma_5 iz \overleftrightarrow{D} u(0) \right) \gamma^\mu \slashed{z} d(0)  |p\rangle &= -\lambda_1 f_1^u (pz)m_p  \slashed{z} N(p),
\end{align}
where the first three leading conformal-spin coupling constants $f_N$, $\lambda_1$, and $\lambda_2$ are given by the leading-order local operator matrix elements, whilst the remaining five parameters, associated with one-derivative operators, correspond to next-to-leading conformal-spin contributions.
Their values  at scale $\mu=1$ GeV  have been obtained from QCD sum rule analyses~\cite{Braun:2000kw}, and we collect them as  
\begin{align} 
f_N&=(5.3\pm 0.5)\times 10^{-3} \text{GeV}^2,\,\lambda_1 =-(2.7\pm 0.9)\times 10^{-2} \text{GeV}^2,\,\lambda_2 =-(5.1\pm 1.9)\times 10^{-2} \text{GeV}^2,  \nonumber\\
V_1^d&= 0.23\pm 0.03, \,A_1^u=0.38\pm 0.15,\, f_1^d=0.6\pm 0.2,\, f_2^d=0.15\pm 0.06, \,f_1^u=0.22\pm 0.15.
\end{align} 
It is evident that these parameters suffer from poor constraints, with uncertainties for certain parameters as large as 50\%, which represent the primary source of theoretical uncertainty in our predictions. As will be further demonstrated in the next section, $\lambda_1$ gives rise to the dominant theoretical error, due to both its larger relative uncertainty and its coupling to the most numerous matrix elements, as detailed in Eq.~(\ref{eq:fn}).

At the quark level, the effective Hamiltonian governing the $b\to c\bar u d$ transition is given by~\cite{Buchalla:1995vs}
\begin{eqnarray}
\mathcal{H}_{\mathrm{eff}}&=&\frac{G_F}{\sqrt{2}}V_{cb}V^*_{ud}\left[C_1(\mu)O_1(\mu)+C_2(\mu)O_2(\mu)\right] +\mathrm{H.c.},
\end{eqnarray}
where $G_F$ is the Fermi constant, $V_{cb}$ and $V_{ud}$ are CKM matrix elements, and $C_{1,2}(\mu)$ are the Wilson coefficients evaluated at the renormalization scale $\mu$. The four-quark operators $O_{1,2}$ are defined as
\begin{eqnarray}
O_1&=& \bar{c}_i \gamma_\mu(1-\gamma_5) b_j  \otimes \bar{d}_j  \gamma^\mu(1-\gamma_5) u_i, \nonumber\\
O_2&=& \bar{c}_i \gamma_\mu(1-\gamma_5) b_i \otimes \bar{d}_j  \gamma^\mu(1-\gamma_5) u_j.
\end{eqnarray}
The corresponding expressions for the $b\to u\bar c d$ transition can be obtained by an obvious interchange of quark flavors. 
In both cases, penguin operators do not contribute due to their flavor structures. Since the $B$ meson is a pseudoscalar state and the two final-state baryons both carry spin $1/2$, the relative orbital angular momentum between them can be in $S$- and $P$-wave configurations. The decay amplitude is then obtained by sandwiching the effective Hamiltonian between the initial and final hadronic states,
\begin{eqnarray}\label{eq:amp}
\mathcal{M}=\langle \Lambda_c^+\bar p|\mathcal{H}_{\mathrm{eff}}| \bar B^0\rangle=\bar u \left(A+B\gamma_5\right)v,
\end{eqnarray}
where $A$ and $B$ correspond to the scalar and pseudoscalar structures, respectively. The $S$- and $P$-wave amplitudes are given by $S=\sqrt{Q_-}B$ and $P=\sqrt{Q_+}A$, respectively, with $Q_{\pm}=M^2-(m_{\Lambda_c}\pm m_p)^2$. The coefficients $A$ and $B$ can be expressed in a factorized form as
\begin{eqnarray}\label{eq:ab}
A/B=\frac{2\sqrt{2}f_{\Lambda_c}   \pi^2 G_FV_{cb}V^*_{ud}}{27\sqrt{3}}\sum_{R_{ij}}
\int\mathcal{D}x\mathcal{D}b
\alpha_s^2(t_{R_{ij}})e^{-S_B-S_{\Lambda_c}-S_p}\Omega_{R_{ij}}(b_l,b'_l,b_q) a_{R_{ij}} H^{ A/B}_{R_{ij}}(x_l,x'_l,y),
\end{eqnarray}
with the integration measures
\begin{eqnarray}
\mathcal{D}x&=&dx_1dx_2dx_3\delta(1-x_1-x_2-x_3)dx'_1dx'_2dx'_3\delta(1-x'_1-x'_2-x'_3)dy,\nonumber\\
\mathcal{D}b&=& d^2\textbf{b}_qd^2\textbf{b}_2d^2\textbf{b}_3d^2\textbf{b}'_2d^2\textbf{b}'_3.
\end{eqnarray}
The $\delta$ functions enforce momentum conservation among the valence quarks. 
 The summation in Eq.~(\ref{eq:ab}) runs over all possible Feynman diagrams, denoted collectively as $R_{ij}$. Here, $R = C$ and $R = E$ label the $W$-emission and $W$-exchange topologies, respectively, as shown in Figs.~\ref{fig:C} and~\ref{fig:E}. The subscript $i = a,b,c,d,e,f,g$ specifies the diagram type, while $j\in [1,7]$ distinguishes the specific gluon-attachment configuration. 
The coefficient $a_{R_{ij}}$ denotes the product of CKM matrix elements and Wilson coefficients associated with a given diagram $R_{ij}$. The function $H_{R_{ij}}$ represents the hard kernel, encoding the spinor and Dirac structure of the short-distance interaction. The factor $\Omega_{R_{ij}}(b_l,b'_l,b_q)$ arises from the Fourier transformation of virtual quark and gluon propagators in transverse-momentum space. The variables $b_l$, $b'_l$, and $b_q$ are conjugate to the parton transverse momenta $k_{lT}$, $k'_{lT}$, and $q_T$, respectively. Explicit expressions for $\Omega_{R_{ij}}$ for both $W$-emission and $W$-exchange topologies can be found in Ref.~\cite{Rui:2024xgc}.

The hard scales $t_{R_{ij}}$ are chosen as the maximal virtuality of internal propagators and factorization scales,
\begin{eqnarray}\label{eq:tt}
t_{R_{ij}}=\max(\sqrt{|t_A|},\sqrt{|t_B|},\sqrt{|t_C|},\sqrt{|t_D|},w,w',1/b_q),
\end{eqnarray}
where $t_{A,B}$ correspond to two gluon propagators, while $t_{C,D}$ are associated with two quark propagators. The infrared cutoff parameters $w$ and $w'$ are defined as the minimal inverse transverse separations among the valence quarks in the baryons~\cite{Shih:1998pb}.  The choice in Eq.~(\ref{eq:tt}) implies the typical hard scales for the decays under study are of order \(t\sim \sqrt{\Lambda_{\text{QCD}}M}\) with $\Lambda_{\text{QCD}}$ being the QCD scale, corresponding to  $\alpha_s(t) \sim 0.39$.  As pointed out in~\cite{Han:2022srw}, an additional constraint is imposed on the hard scale, requiring $t_{R_{ij}} \ge 1\,\text{GeV}$. This lower bound ensures that the scale does not fall below $1\,\text{GeV}$, at which the baryon  LCDAs are defined.
Consequently, all contributions originate from the region where $\alpha_s/\pi \le 0.157$, indicating that the perturbative expansion remains reasonably under control and that the application of perturbative QCD to these decays is feasible.   Our numerical results show that both the branching ratios and the asymmetry parameters are  sensitive to this lower constraint, implying that the maximal virtuality of internal propagators and the factorization scales are insufficient to ensure that the energy scale lies within the perturbative regime.
Therefore, retaining this constraint is essential, as it avoids the double counting of certain nonperturbative contributions in the hard scattering kernel. 
The quantities specific to each diagram, such as $a_{R_{ij}}$, $H_{R_{ij}}$, and $t_{R_{ij}}$, are collected in the Appendix.

The Sudakov exponents associated with the initial and final states are given by~\cite{Ali:2007ff,He:2006vz,Kundu:1998gv}
\begin{align}\label{eq:sud}
S_{B}&=s(q^-_1,b_q)+\frac{5}{3}\int^t_{1/b_q}d \bar \mu \frac{\gamma(\alpha_s(\bar \mu))}{\bar \mu},\nonumber\\
S_{\Lambda_c}&=s_c(k_1^{\prime-},cw^\prime)+\sum_{l=2}^3s(k_l^{\prime-},cw^\prime)+\frac{8}{3}\int^t_{cw^\prime}d \bar \mu \frac{\gamma(\alpha_s(\bar \mu))}{\bar \mu},\nonumber\\
S_{p}&=\sum_{l=1}^3s(k_l^{+},cw)+3\int^t_{cw}d \bar \mu \frac{\gamma(\alpha_s(\bar \mu))}{\bar \mu},
\end{align}
where the quark anomalous dimension is $\gamma(\alpha_s(\bar \mu)) = -\alpha_s(\bar \mu)/\pi$. The integrals involving $\gamma(\alpha_s)$ represent single-logarithmic renormalization-group evolution from the hard scale $t$ down to the factorization scales. The functions $s$ and $s_c$ describe double-logarithmic Sudakov resummation for light and charm quarks, respectively, and their next-to-leading-logarithmic expressions can be found in Refs.~\cite{Shih:1998pb,Ali:2007ff,Liu:2023kxr,Liu:2020upy}. For numerical consistency, a two-loop running coupling is used, with $\Lambda_{\mathrm{QCD}}=0.225$ GeV for $n_f=5$, determined from the world-average value $\alpha_s(M_Z)=0.1179$~\cite{ParticleDataGroup:2026aaa}. The parameter $c$ is introduced to parametrize residual resummation uncertainties, with $c=1.05$ for the $\Lambda_c$ baryon~\cite{Rui:2024xgc} and $c=1.14$ for the proton~\cite{Kundu:1998gv}.

\section{Numerical results}\label{sec:results}

For the numerical analysis, we employ the following input parameters~\cite{ParticleDataGroup:2026aaa, Wang:2010fq,He:2006vz,Rui:2021kbn}:
\begin{eqnarray}
\lambda &=&0.22650, \quad A=0.790,  \quad \bar{\rho}=0.141, \quad \bar{\eta}=0.357,\nonumber\\
M_{B}&=&5.28 \text{GeV}, \quad m_{p}=0.938\text{GeV}, \quad m_{\Lambda_c}=2.286\text{GeV}, \quad m_b=4.8\text{GeV},\nonumber\\
f_{B}&=&0.19 \text{GeV}, \quad     f_{\Lambda_c}=0.022  \text{GeV}^3, \quad \tau_{B}=1.52 \text{ps}.
\end{eqnarray}
The remaining nonperturbative parameters entering the hadronic LCDAs have been specified in the previous section.

The branching fractions and angular asymmetry parameters are defined as~\cite{Geng:2023nia,Wang:2024qff}
\begin{eqnarray}\label{eq:two}
\mathcal{B}&=&\frac{P_c\tau_B}{8\pi M^2}|\mathcal{M}|^2=\frac{P_c\tau_B}{4\pi M^2}(|H_+|^2+|H_-|^2),  \nonumber\\
\alpha&=&\frac{|H_+|^2-|H_-|^2}{|H_+|^2+|H_-|^2},  \nonumber\\
\beta&=&\frac{2Im(H_+H_-^*)}{|H_+|^2+|H_-|^2},  \nonumber\\
\gamma&=&\frac{2Re(H_+H_-^*)}{|H_+|^2+|H_-|^2},
\end{eqnarray}
where $P_c=\sqrt{Q_+Q_-}/(2M)$ denotes the magnitude of the three-momentum of either final-state baryon in the $B$-meson rest frame. The helicity amplitudes $H_\pm$ are related to the invariant amplitudes introduced in Eq.~(\ref{eq:amp}) through
\begin{eqnarray}
H_{\pm}=\frac{1}{\sqrt{2}}\left(\sqrt{Q_+}A\mp\sqrt{Q_-}B\right).
\end{eqnarray}
The parameter $\alpha$ characterizes the up-down asymmetry in the baryon angular distribution, while $\beta$ and $\gamma$ encode the polarization information of the final-state baryons. These observables satisfy the normalization condition
\begin{eqnarray}
\alpha^2+\beta^2+\gamma^2=1.
\end{eqnarray}

In the numerical analysis, the branching fractions are obtained by coherently summing all contributing amplitudes, including both the $W$-emission and $W$-exchange topologies. The resulting predictions for the decay $\bar B^0\to\Lambda_c^+\bar p$ based on different models of the $\Lambda_c$ LCDAs are
\begin{eqnarray}\label{eq:various} 
\mathcal{B}(\bar {B}^0\rightarrow\Lambda_c^+\bar p)= \left\{ \begin{array}{l} 
(1.64^{+0.58+0.06+0.59+0.46}_{-0.42-0.10-0.43-0.24})\times 10^{-5}, \quad \text{exponential model}, \\ (1.64^{+0.64+0.08+0.53+0.51}_{-0.42-0.00-0.44-0.24})\times 10^{-5}, \quad \text{QCDSR model},\\ (2.08^{+0.61+0.00+0.66+0.46}_{-0.52-0.07-0.62-0.32})\times 10^{-5}, \quad \text{Gegenbauer model}.\\ 
\end{array} 
\right. 
\end{eqnarray}
The four quoted uncertainties originate from different sources. The first arises from the shape parameter $\omega_b=0.40$ GeV in the $B$-meson wave function~\cite{Hua:2020usv}. Varying $\omega_b$ by $\pm0.05$ GeV leads to an uncertainty of roughly $30\%$ in the branching fraction, making it one of the dominant sources of theoretical error. The second uncertainty is associated with the shape parameters entering the various models of the $\Lambda_c$ LCDAs. As discussed in the previous section, the improved parametrizations adopted in this work significantly reduce the sensitivity of the predictions to the poorly known nonperturbative structure of the $\Lambda_c$ baryon. Indeed, the central values obtained from the exponential and QCDSR models are almost identical, while the prediction based on the Gegenbauer model differs only moderately, indicating a reasonable stability of the results with respect to the choice of the $\Lambda_c$ LCDAs. The third uncertainty originates from the proton LCDAs. It is estimated by varying the nonperturbative parameters associated with the leading conformal-spin operators and adding the resulting errors in quadrature. Specifically, we use $f_N=(0.53\pm0.05)\times10^{-2}~\mathrm{GeV}^2$, $\lambda_1=(-2.7\pm0.9)\times10^{-2}~\mathrm{GeV}^2$, and $\lambda_2=(5.1\pm1.9)\times10^{-2}~\mathrm{GeV}^2$~\cite{Braun:2000kw,Han:2022srw}. Among these inputs, the higher-twist parameter $\lambda_1$ provides the dominant contribution to the uncertainty and therefore deserves further investigation in future studies. The last uncertainty reflects the residual renormalization-scale dependence of the leading-order PQCD calculation. To estimate the potential size of higher-order QCD corrections, the hard scale $t$ is varied within the range $0.8t$--$1.2t$. The resulting uncertainty is moderate, suggesting that the perturbative expansion remains under reasonable control. A more precise assessment, however, requires a complete next-to-leading-order analysis in $\alpha_s$. Combining all sources of uncertainty, we find that the dominant theoretical errors are still associated with nonperturbative hadronic inputs, particularly those entering the $B$-meson wave function and the higher-twist proton LCDAs. Nevertheless, all three models of the $\Lambda_c$ LCDAs lead to branching fractions of order $10^{-5}$, in good agreement with the measured value. This consistency indicates that the PQCD framework, supplemented by the improved treatment of higher-twist effects and the inclusion of both $W$-emission and $W$-exchange contributions, provides a realistic description of the decay $\bar B^0\to\Lambda_c^+\bar p$.

The predicted branching fractions are of order $10^{-5}$, with central values ranging from $1.64\times10^{-5}$ to $2.08\times10^{-5}$ depending on the choice of the $\Lambda_c$ LCDA model. For comparison, the experimental world average reported by the PDG~\cite{ParticleDataGroup:2026aaa}, together with representative theoretical predictions available in the literature~\cite{Hsiao:2019wyd,He:2006vz,Cheng:2002sa,Cheng:2001ub,Jarfi:1990ej,Ball:1990fw,Chernyak:1990ag}, are collected in Table~\ref{tab:ratio}. The current world-average value, $\mathcal{B}(\bar B^0\to\Lambda_c^+\bar p)=(1.52\pm0.17)\times10^{-5}$, is obtained from the measurements performed by Belle~\cite{Belle:2002gir} and BABAR~\cite{BaBar:2008get}. It can be seen that the predictions based on the Exponential and QCDSR models are in excellent agreement with the experimental result. In the following, we therefore adopt the Exponential model as the default choice.  It is worth noting that early studies based on the pole model~\cite{Jarfi:1990ej}, the diquark model~\cite{Ball:1990fw}, and QCD sum rules~\cite{Chernyak:1990ag} typically predicted branching fractions at the $10^{-3}$ level, exceeding the measured value by nearly two orders of magnitude. Such large predictions are incompatible with current experimental data. The upper limits estimated in Refs.~\cite{Cheng:2002sa,Cheng:2001ub} are consistent with the observed branching fraction, although they remain subject to sizable model uncertainties. The result obtained in Ref.~\cite{Hsiao:2019wyd} is somewhat lower than the experimental central value, which can be attributed to the fact that only the $W$-exchange contribution was taken into account in that analysis. In baryonic $B$ decays, the color structure differs significantly from that of mesonic decays. Owing to the totally antisymmetric color wave function of the baryon, the internal $W$-emission topology is not necessarily subject to the conventional color suppression encountered in mesonic $B$ decays. As will be demonstrated below, the internal $W$-emission contribution provides the dominant component of the $\bar B^0\to\Lambda_c^+\bar p$ decay amplitude. The PQCD analysis of Ref.~\cite{He:2006vz} predicts a branching fraction in the range $(2.3$--$5.1)\times10^{-5}$, which is significantly larger than the experimental measurement. This discrepancy can be traced to the treatment of the $W$-exchange topology. In Ref.~\cite{He:2006vz}, the $W$-exchange contribution was assumed to be helicity suppressed, by analogy with the annihilation amplitude in purely leptonic $B$ decays~\cite{BaBar:2014omp}, and was therefore neglected. Our analysis indicates that this approximation is not appropriate for charmful baryonic decays. Numerically, the $W$-exchange amplitude amounts to roughly one half of the internal $W$-emission amplitude and interferes destructively with it. As a consequence, including the $W$-exchange contribution substantially reduces the predicted branching fraction, leading to much better agreement with the experimental data. This finding provides further evidence that helicity suppression is considerably alleviated in charmful baryonic decays owing to the presence of the heavy charm quark~\cite{Hsiao:2019wyd}.

\begin{table}[!htbh]
\caption{
The comparison of theoretical predictions for branching ratio of the $\bar {B}^0\rightarrow\Lambda_c^+\bar p$ decay
in different methods and the latest measured experimental values. Errors are explained in the text.}\label{tab:ratio}
	\begin{tabular}[t]{lcccccc}
	\hline\hline
    &PDG~\cite{ParticleDataGroup:2026aaa}  & \cite{Hsiao:2019wyd}& \cite{He:2006vz}  &\cite{Cheng:2002sa,Cheng:2001ub} &\cite{Jarfi:1990ej,Ball:1990fw,Chernyak:1990ag}\\ \hline
$\mathcal{B}( \bar{B}^0\rightarrow \Lambda_c^+ \bar p)(10^{-5})$
&\quad $1.52\pm0.17 $ &\quad $1.0^{+0.4}_{-0.3}$ &\quad$2.3-5.1$ &\quad$\leq(1.1-3.2)$ &\quad$\sim100$\\
		\hline\hline
	\end{tabular}
\end{table}

\begin{table}[!htbh]
	\caption{Contributions from the $W$-emission ($C$) and $W$-exchange ($E$) topologies to the decay amplitudes of the $\bar {B}^0\rightarrow\Lambda_c^+\bar p$ decay. The last two columns are their relative magnitude and phase (in radian), respectively.}\label{tab:CE} 
	\begin{tabular}[t]{lcccc}
	\hline\hline
Amplitude  &\quad$C$\quad & \quad$E$\quad &\quad$|\frac{E}{C}|$\quad & \quad$\text{Arg}(E/C)$\quad\\ \hline
$S(10^{-8}\text{GeV})$ &$-0.9+3.3 \textit{i}$  &$0.3-0.8\textit{i}$    &0.24 &-3.05\\
$P(10^{-8}\text{GeV})$ &$-0.1-2.1\textit{i}$    &$0.6-0.2\textit{i}$  &0.30 &1.26\\
$ |\mathcal{M}|(10^{-8}\text{GeV})$ &$4.0$ &$1.0$ &0.25&$\cdots$\\
$ \mathcal{B} $ &$2.2\times10^{-5}$ &$1.5\times10^{-6}$ &$\cdots$&$\cdots$\\
		\hline\hline
	\end{tabular}
\end{table}

To illustrate the relative importance of the two topologies, the partial-wave amplitudes arising from internal $W$ emission ($C$) and $W$ exchange ($E$) are compared in Table~\ref{tab:CE}.~\footnote{Unless otherwise stated, only the Exponential model for the $\Lambda_c$ LCDAs is considered in the subsequent analysis. Explicit calculations show that the other LCDA models lead to qualitatively similar results.} The relative magnitudes and phases are listed in the last two columns. The decay is clearly dominated by the internal $W$-emission contribution, while the $W$-exchange topology remains sizable and cannot be neglected. In particular, the ratio $|E/C|$ reaches about $0.3$ for the $P$-wave amplitudes. Table.~\ref{tab:CE} further shows that the $C$ and $E$ amplitudes exhibit opposite interference patterns in different partial waves. The relative phases indicate destructive interference in the $S$ wave and mildly constructive interference in the $P$ wave. Overall, however, the net interference between the $C$ and $E$ topologies is destructive, leading to a reduced branching fraction for $\bar{B}^0\to\Lambda_c^+\bar p$ and consequently improving the agreement with experiment. A similar destructive interference pattern between internal $W$ emission and $W$ exchange was previously observed in our study of $\bar{B}^0\to\Lambda_c^+\bar{\Lambda}_c^-$~\cite{Rui:2024xgc}, and was subsequently supported by the recent LHCb measurement~\cite{LHCb:2025ueu}.

When the $W$-exchange contribution is switched off, we obtain a central branching fraction of $2.2\times10^{-5}$ (see Table.~\ref{tab:CE}), which is slightly smaller than the PQCD prediction of $(2.3$--$5.1)\times10^{-5}$ reported in Ref.~\cite{He:2006vz}. This difference can be mainly attributed to the inclusion of higher-twist contributions from the nonperturbative hadronic LCDAs, whose impact will be discussed in detail below. Conversely, retaining only the $W$-exchange topology yields a branching fraction of $1.5\times10^{-6}$, which is approximately a factor of seven smaller than the central value, $(1.0^{+0.4}_{-0.3})\times10^{-5}$, obtained in Ref.~\cite{Hsiao:2019wyd}. Although the individual contributions differ substantially from the results of Refs.~\cite{Hsiao:2019wyd,He:2006vz}, the combined effects of the internal $W$-emission and $W$-exchange topologies lead to a branching fraction in much better agreement with the experimental measurement.

As noted above, our analysis incorporates several higher-twist effects that were neglected or only partially included in previous studies. These include the subleading LCDA $\bar{\phi}_B$ of the $B$ meson, the $\Lambda_c$ LCDAs up to twist 4, and the proton LCDAs up to twist 6. Numerically, these contributions are found to play a non-negligible role. To quantify their impact, we decompose the corresponding decay amplitudes into their $S$- and $P$-wave components and present the real and imaginary parts separately. This decomposition allows us to identify the interference patterns generated by different LCDA components and to assess their relative importance. We now discuss these contributions in turn.

We first examine the contributions from the leading-twist LCDA $\phi_B$ and the subleading LCDA $\bar{\phi}_B$ of the $B$ meson, whose numerical results are summarized in Table~\ref{tab:sub}. A notable feature is that $\bar{\phi}_B$ generates a sizable imaginary part in the $P$-wave amplitude, making its contribution to the total amplitude magnitude $|\mathcal M|$ comparable to that of the leading component. As shown in the last row of Table~\ref{tab:sub}, the contribution from $\bar{\phi}_B$ reaches roughly one third of that from $\phi_B$, indicating that the subleading LCDA effect is numerically comparable to a typical next-to-leading-order correction. Furthermore, the contributions from $\phi_B$ and $\bar{\phi}_B$ interfere constructively. As a result, including the $\bar{\phi}_B$ term enhances the magnitude of the total decay amplitude by approximately $30\%$. These results demonstrate that the subleading LCDA $\bar{\phi}_B$ provides a significant correction and should be retained in a realistic description of the $\bar{B}^0\to\Lambda_c^+\bar p$ decay.

\begin{table}[htbh]
	\caption{Partial-wave amplitudes and total amplitude magnitudes (in units of $10^{-8}\mathrm{GeV}$) from the leading-twist and subleading-twist $B$-meson LCDAs, $\phi_B$ and $\bar{\phi}_B$. The last column gives their combined contribution. }
	\label{tab:sub} 
	\begin{tabular}[t]{lccc}
	\hline\hline
 Amplitude ($10^{-8}\text{GeV}$)                         &$\phi_B$ & $\bar \phi_B$ & $\phi_B+\bar \phi_B$ \\ \hline
 $S$                    & $-0.4+2.2\textit{i}$ &  $-0.2+0.3\textit{i}$ & $-0.6+2.5\textit{i}$ \\
 $P$                    & $0.5-1.5\textit{i}$ &  $-0.1-0.8\textit{i}$ & $0.4-2.3\textit{i}$ \\
$ |\mathcal{M}|$       & $2.7 \times 10^{-8}$& $0.9 \times 10^{-8}$& $3.5 \times 10^{-8}$ \\
		\hline\hline
	\end{tabular}
\end{table}	

In Table~\ref{tab:twsit}, we present the partial-wave amplitudes arising from different combinations of the $\Lambda_c$ and proton LCDAs, including contributions from both the $\phi_B$ and $\bar{\phi}_B$ components of the $B$-meson LCDA. The rows (columns) correspond to the twists of the $\Lambda_c$ (proton) LCDAs. A clear hierarchical structure emerges from the numerical results. For fixed $\Lambda_c$ twist, the magnitude of the amplitude decreases systematically with increasing proton twist. Contributions involving twist-3, twist-4, and twist-5 proton LCDAs are typically of order $10^{-8}\mathrm{GeV}$, whereas those associated with twist-6 LCDAs are further suppressed to the level of $10^{-9}\mathrm{GeV}$. On the $\Lambda_c$ side, the twist-3 LCDA provides the dominant contribution, followed by the twist-4 component, while the leading-twist contribution is comparatively small. Consequently, the largest contribution to the decay amplitude originates from the convolution of the twist-3 $\Lambda_c$ and twist-3 proton LCDAs. In contrast, amplitudes involving the leading-twist $\Lambda_c$ LCDA are generally suppressed by approximately one order of magnitude. This hierarchy reflects the well-known endpoint-enhanced behavior of higher-twist heavy-baryon LCDAs, which can compensate for their nominal power suppression, as discussed in Ref.~\cite{Han:2022srw}. The sizable higher-twist effects observed here are therefore not unexpected. At the same time, the numerically smallest contributions arise from the combination of twist-4 $\Lambda_c$ and twist-6 proton LCDAs. These terms are suppressed by the factor $(m_p/M)^3$ and contribute only at the level of $\mathcal{O}(10^{-10})\mathrm{GeV}$. The strong suppression of the highest-twist configurations, together with the observed hierarchical pattern, provides nontrivial evidence for the convergence and self-consistency of the baryonic twist expansion employed in the present PQCD framework. The interference pattern among different twist components is also noteworthy. We find that the twist-2 and twist-3 contributions of the $\Lambda_c$ LCDA interfere predominantly destructively, whereas the twist-2 and twist-4 contributions tend to interfere constructively. These features originate from the distinct Lorentz structures and Dirac projections entering the definitions of the corresponding LCDAs. Such interference effects play an important role in shaping the final decay amplitude and contribute to the observed hierarchy among the various twist sectors.

\begin{table}[htbh]
	\caption{Partial-wave amplitudes and total amplitude magnitudes (in units of $10^{-9}\mathrm{GeV}$) arising from different twist combinations of the $\Lambda_c$ and proton LCDAs. The last column (row) gives the sum over all proton ($\Lambda_c$) twist contributions.}
	\label{tab:twsit} 
	\begin{tabular}[t]{lcccccc}
	\hline\hline
     & Amplitude    & Twist-3                 & Twist-4                 & Twist-5           & Twist-6          &Total  \\ \hline
    Twist-2 & $S$ & $1.4 +0.2\textit{i} $  &$1.6-4.3\textit{i}$    &$3.0+1.2\textit{i}$   &$-0.4+0.7\textit{i}$  &$5.5-2.0\textit{i}$ \\
            & $P$ & $-0.1 +5.4\textit{i} $    &$1.2-0.4\textit{i}$    &$0.9 -0.4\textit{i}$   &$-0.5+0.6\textit{i}$  &$1.5+5.1\textit{i}$ \\
   &$ |\mathcal{M}|$ & $5.6$    & $4.8$    & $3.3$    & $1.1$   & $7.9$   \\
\hline 
   Twist-3  & $S$   & $-11.8 +28.3\textit{i} $   &$-3.1+20.0\textit{i}$     &$2.4-7.0\textit{i}$   &$0.0+1.2\textit{i}$  &$-12.2+42.6\textit{i}$ \\
            & $P$   & $3.7-9.4\textit{i} $  &$-15.0-2.0\textit{i}$      &$3.0 -6.5\textit{i}$   &$0.0+1.1\textit{i}$  &$-8.4-16.7\textit{i}$ \\
   &$ |\mathcal{M}|$  & $32.3$    & $25.3$    & $10.3$    & $1.6$   & $48.1$   \\
\hline 
   Twist-4 & $S$   & $-3.9 -9.1\textit{i} $  &$3.1-4.9\textit{i}$    &$1.5-1.5\textit{i}$   &$-0.3+0.4\textit{i}$  &$0.7-15.2\textit{i}$ \\
            & $P$  & $5.7 -6.3\textit{i} $  &$5.6-5.0\textit{i}$    &$0.2 -0.5\textit{i}$    &$-0.1+0.3\textit{i}$  &$11.3-11.5\textit{i}$ \\
   &$ |\mathcal{M}|$ & $13.0$    & $9.5$    & $2.2$    & $0.6$   & $22.2$   \\
\hline 
   Total   & $S$   & $-14.2+19.3\textit{i} $    &$2.0+11.2\textit{i}$   &$6.8-7.3\textit{i}$   &$-0.7+2.2\textit{i}$   &$-6.1+25.3\textit{i}$ \\
           & $P$   & $9.3-10.3\textit{i} $   &$-8.1-7.5\textit{i}$     &$4.0 -7.4\textit{i}$   &$-0.6+2.1\textit{i}$   &$4.4-23.1\textit{i}$ \\
   &$ |\mathcal{M}|$ & $27.7$    & $15.8$    & $13.1$    & $3.2$   & $35.1$   \\   
		\hline\hline
	\end{tabular}
\end{table}	

With the decay amplitudes determined, the asymmetry parameters can be obtained directly from Eq.~(\ref{eq:two}). The resulting PQCD predictions based on the three $\Lambda_c$ LCDA models are presented in Table~\ref{tab:asys}, where the quoted uncertainties are evaluated from the same input variations used throughout this work.  Since the angular observables are proportional to the ratio of helicity amplitudes, the normalization uncertainties associated with $\omega_b$ and the scale $t$ largely cancel in these ratios. However, $\lambda_1$ enters to proton LCDAs of different twists with distinct coefficients and signs, giving rise to twist-dependent interference patterns that affect $H_+$ and $H_-$ asymmetrically. As a result, these effects do not cancel in the ratios.  The dominant theoretical uncertainty originates from the nonperturbative parameter $\lambda_1$ in the proton LCDA, highlighting the importance of improved determinations of baryonic distribution amplitudes for future precision studies. As shown in Table~\ref{tab:asys}, all three LCDA models predict a large and positive up--down asymmetry parameter $\alpha$, with the Exponential and QCDSR models yielding values close to unity. This feature can be traced to the strong cancellation between two partial-wave amplitudes of comparable magnitude and opposite sign, as illustrated in Table~\ref{tab:sub}. Consequently, the asymmetry parameters $\beta$ and $\gamma$ remain comparatively small. The emergence of a large positive $\alpha$ accompanied by suppressed values of $\beta$ and $\gamma$ constitutes a distinctive prediction of the present PQCD framework and provides a potentially sensitive probe of the underlying decay dynamics. Despite differences in the modeling of the $\Lambda_c$ LCDA, the predicted asymmetry parameters are fully compatible within theoretical uncertainties, demonstrating the robustness of the qualitative pattern.

\begin{table}[!htbh]
	\caption{Angular distribution parameters with three different LCDA models of $\Lambda_c$ baryon (Exponential, QCDSR, and Gegenbauer models) for $\bar B^0 \rightarrow \Lambda_c^+ \bar p$ decay. The theoretical uncertainties are the same as for branching ratios in Eq.~(\ref{eq:various}).}
	\label{tab:asys}
	\begin{tabular}[t]{lccc}
	\hline\hline
Model  
&$\alpha$ 
& $\beta$ 
&$\gamma$ \\ 
\hline
Exponential  
&$0.993^{+0.003+0.000+0.009+0.003}_{-0.005-0.008-0.064-0.002}$
&$0.047^{+0.041+0.025+0.030+0.000}_{-0.054-0.087-0.059-0.042}$
&$-0.104^{+0.024+0.012+0.223+0.016}_{-0.020-0.062-0.301-0.021}$\\
QCDSR        
&$0.993^{+0.005+0.003+0.006+0.004}_{-0.005-0.003-0.038-0.001}$
&$0.100^{+0.041+0.000+0.033+0.000}_{-0.068-0.028-0.069-0.031}$
&$-0.068^{+0.012+0.005+0.227+0.016}_{-0.005-0.039-0.253-0.015}$\\
Gegenbauer 
&$0.933^{+0.009+0.001+0.014+0.006}_{-0.031-0.035-0.074-0.014}$
&$-0.310^{+0.058+0.071+0.119+0.003}_{-0.095-0.116-0.144-0.025}$
&$-0.180^{+0.035+0.077+0.250+0.025}_{-0.038-0.085-0.253-0.028}$\\
\hline
\hline
	\end{tabular}
\end{table}

Finally, we investigate the highly suppressed decay $\bar B^0 \to \bar\Lambda_c^- p$, which proceeds via the quark-level transition $b \to u \bar c d$. Relative to the favored mode $\bar B^0 \to \Lambda_c^+ \bar p$, this channel is doubly CKM suppressed by the factors $|V_{ub}/V_{cb}|$ and $|V_{cd}/V_{ud}|$, while the relative weak phase between the two amplitudes is given by the CKM angle $\gamma$, analogous to the case of the two amplitudes of $\Lambda_b \to D^0 \Lambda$ and $\Lambda_b \to \bar{D}^0\Lambda$ decays~\cite{Rui:2026ihu}. In principle, time-dependent measurements of these decays in the neutral-$B$ system could provide access to $\gamma$ through the interference between $B^0$--$\bar B^0$ mixing and decay amplitudes. The numerical results for $\bar B^0 \to \bar\Lambda_c^- p$, obtained using the Gegenbauer, QCDSR, and Exponential models for the $\Lambda_c$ LCDAs, are summarized in Table~\ref{tab:asys2}. The three models yield consistent predictions within uncertainties, which are dominated by the poorly constrained proton LCDA parameters. The predicted branching fractions are of order $10^{-8}$, with central values ranging from $1.1\times10^{-8}$ to $1.6\times10^{-8}$. This magnitude is comparable to that of the purely baryonic decay $\bar B^0\to p\bar p$, for which LHCb measured $\mathcal{B}(\bar B^0\to p\bar p)=(1.27\pm0.15\pm0.05\pm0.04)\times10^{-8}$~\cite{LHCb:2022oyl}. The decay $\bar B^0 \to \bar\Lambda_c^- p$ may therefore become accessible in future experimental analyses. Combining the branching fraction predictions for the suppressed and favored channels, we obtain
\begin{eqnarray}
\mathcal{R}
=\left|
\frac{A(\bar B^0\to\bar\Lambda_c^- p)}
{A(\bar B^0\to\Lambda_c^+\bar p)}
\right|
\approx (2\text{--}3)\%.
\end{eqnarray}
The smallness of this amplitude ratio indicates that the sensitivity to the weak phase $\gamma$ is expected to be rather limited in this decay system. For the angular observables, all three LCDA models lead to a common qualitative pattern: $\beta$ and $\gamma$ are negative, with typical values in the ranges $[-0.8,-0.6]$ and $[-0.4,-0.2]$, respectively, whereas $\alpha$ remains positive, with central values between $0.400$ and $0.732$. The theoretical uncertainties are again dominated by the proton LCDA parameters, which presently limit the precision of these predictions. Improved determinations of baryonic LCDAs from nonperturbative approaches would therefore be highly valuable for sharpening the phenomenological predictions. The angular observables presented here provide additional targets for future experimental studies of heavy-flavor baryonic decays.

\begin{table}[!htbh]
	\caption{ Same as Table~\ref{tab:asys} but for $\bar B^0 \rightarrow \bar \Lambda_c^- p$ decay. }
	\label{tab:asys2}
	\begin{tabular}[t]{lcccc}
	\hline\hline
Model 
&$\mathcal{B}(10^{-8})$ 
&$\alpha$ 
& $\beta$ 
&$\gamma$ \\ 
\hline
Exponential  
&$1.1^{+0.5+0.1+0.9+0.5}_{-0.3-0.1-0.5-0.3}$
&$0.689^{+0.033+0.020+0.220+0.063}_{-0.040-0.000-0.461-0.042}$
&$-0.700^{+0.041+0.022+0.271+0.066}_{-0.046-0.002-0.155-0.043}$
&$-0.183^{+0.037+0.015+0.368+0.015}_{-0.023-0.005-0.411-0.000}$\\
QCDSR        
&$1.6^{+0.6+0.0+0.9+0.6}_{-0.5-0.1-0.6-0.4}$ 
&$0.400^{+0.102+0.041+0.362+0.099}_{-0.030-0.000-0.563-0.071}$
&$-0.825^{+0.071+0.014+0.182+0.068}_{-0.032-0.000-0.021-0.045}$
&$-0.399^{+0.040+0.012+0.269+0.032}_{-0.025-0.000-0.250-0.023}$\\
Gegenbauer  
&$1.2^{+0.5+0.2+0.7+0.6}_{-0.3-0.2-0.4-0.3}$ 
&$0.732^{+0.037+0.070+0.170+0.064}_{-0.046-0.106-0.432-0.095}$
&$-0.659^{+0.054+0.093+0.226+0.072}_{-0.044-0.099-0.138-0.086}$
&$-0.172^{+0.000+0.000+0.347+0.029}_{-0.032-0.018-0.515-0.025}$\\
\hline\hline
\end{tabular}
\end{table}

\section{Conclusion}\label{sec:sum}
We have presented a comprehensive PQCD study of the baryonic decay $\bar B^0 \to \Lambda_c^+ \bar p$, incorporating both the $W$-emission and $W$-exchange topologies within a unified framework. Relative to previous PQCD analyses, we systematically include several subleading effects, namely the subleading $B$-meson light-cone distribution amplitude (LCDA), the $\Lambda_c$ LCDAs up to twist-4, and the proton LCDAs up to twist-6. To assess the associated nonperturbative uncertainties, three phenomenological models for the $\Lambda_c$ LCDAs—Gegenbauer, QCDSR, and exponential—are constructed based on heavy-quark symmetry. The resulting predictions are mutually consistent within theoretical uncertainties, indicating a reasonable degree of model stability.

Our analysis shows that the subleading $B$-meson LCDA provides a non-negligible contribution and enhances the decay amplitude significantly. The twist decomposition further reveals that the dominant contribution originates from the combinations of the twist-3 $\Lambda_c$ LCDAs with the twist-3 and twist-4 proton LCDAs, whereas higher-twist terms are progressively suppressed. This hierarchy provides evidence for the convergence of the baryonic LCDA expansion in the present framework. We also find that the inclusion of the $W$-exchange topology reduces the predicted branching fraction compared with earlier PQCD results, leading to substantially improved agreement with the experimental measurement. The dominant theoretical uncertainties arise from nonperturbative parameters entering the baryon LCDAs, while residual hard-scale dependence constitutes a subleading source of error. For branching fractions, the overall uncertainties are typically at the level of $20\%-50\%$, underscoring the importance of more precise nonperturbative inputs. In particular, the angular observables exhibit pronounced sensitivity to the proton LCDA parameters and therefore provide a potentially valuable probe of baryon structure.

As an application of the same framework, we have also investigated the doubly Cabibbo-suppressed decay $\bar B^0 \to \bar\Lambda_c^- p$, whose branching fraction is predicted to be of order $10^{-8}$. 
 The high-luminosity LHCb upgrade, with an integrated luminosity of $300\text{ fb}^{-1}$, would provide an opportunity to observe this suppressed mode and measure its branching fraction with competitive  precision. The ratio between the suppressed and favored amplitudes is found to be only a few percentages, implying limited sensitivity to the CKM angle $\gamma$ through the two channels. Since no penguin contributions are present in either decay mode, direct $CP$ asymmetries vanish in the Standard Model. Consequently, the observation of a sizable direct $CP$ asymmetry would constitute a clean signal of physics beyond the Standard Model.

\section*{ACKNOWLEDGMENT}
This work is supported in part by the National Science Foundation of China under the Grants Nos. 12675124, 12375089, 12435004, and 12075086, and the Natural Science Foundation of Shandong province under the Grant No. ZR2022ZD26 and ZR2022MA035.

\begin{appendix}
\section{FACTORIZATION FORMULAS}\label{sec:for}

\begin{table} [htbh]
\centering
	\caption{The expressions of $a_{R_{ij}}$  for the $W$-emission and $W$-exchange diagrams in the $\bar B^0\rightarrow \Lambda_c^+ \bar p$ decay.} 
	\label{tab:wilson}
	\begin{tabular}[t]{lc}
		\hline\hline
		$R_{ij}$        &$a_{R_{ij}}$                    \\ \hline
		$C_{a1,a2,a4,a5,b1,b4,c1,c3,d1,d3,e1,f1}$        &$C_2-C_1 $                  \\
		
		$C_{a3,b2,c2,f2}$        &$ C_2 +2C_1 $                   \\
		
		$C_{a6,b6,c5,d5,f3}$     &$ 2(C_1-C_2) $               \\
		
		$C_{a7,b7,c6,f4}$        &$ -C_1-2C_2 $               \\
		
		$C_{b3,d2,e2}$           &$ -\frac{1}{4}C_1+C_2 $                \\
		
		$C_{b5}$                 &$\frac{5}{4}C_1+C_2 $               \\
		
		$C_{c4}$                 &$ -C_1-\frac{5}{4}C_2 $                    \\
		
		$C_{c7,d7,e4}$           &$\frac{1}{4}C_2-C_1 $             \\
		
		$C_{d4}$                 &$ \frac{5}{4}(C_1-C_2) $               \\
		
		$C_{d6}$     &$ 4(C_2-C_1) $               \\

        $C_{e3}$     &$ \frac{1}{4}(C_2-C_1) $               \\
		
		$C_{g1}$        &$0 $               \\

        $C_{g2}$        &$ -\frac{9}{4}C_1 $               \\
		
		$C_{g3}$        &$ \frac{9}{4}(C_2-C_1) $                \\
		
		$C_{g4}$        &$ \frac{9}{4}C_2 $                \\
		
		$E_{a1-a5,b1-b5,e1,e2}$        &$ 3C_1+C_2 $              \\
		
		$E_{a6,a7,b6,b7,c1-c4,d1-d4,e3,e4,f3,f4}$        &$ C_2 $           \\
		
		$E_{c5-c7,d5-d7}$        &$C_2-\frac{3}{4}C_1$              \\	
		\hline\hline
	\end{tabular}
\end{table}

\begin{table*} [htbh]	
	\scriptsize
	\centering
	\caption{The virtualities of the internal propagators $t_{A,B,C,D}$  for the $W$-emission diagrams with $\bar x_l^{(')}=1-x_l^{(')}$.}
	\newcommand{\tabincell}[2]{\begin{tabular}{@{}#1@{}}#2\end{tabular}}
	\label{tab:ttt}
	\begin{tabular}[t]{lcccc}
		\hline\hline\
		$R_{ij}$ &  $\frac{t_A}{M^2}$   &$\frac{t_B}{M^2}$  &$\frac{t_C}{M^2}$ &$\frac{t_D}{M^2}$ \\ \hline
		$C_{a1}$    &$f^+ x_1 y$   &$\left(f^--1\right) f^+ x_2 x'_2$
    &$f^+ \left(x_1+x_3\right) y$  &$f^+ \left(\left(f^--1\right) x'_2+y\right)$ \\	

    $C_{a2}$    &$f^+ x_1 y$   &$\left(f^--1\right) f^+ x_2 x'_2$
    &$\left(f^--1\right) f^+ \bar x_1 x'_2$  &$f^+  \left(\left(f^--1\right) x'_2+y\right)$ \\	

    $C_{a3}$    &$f^+ x_1 y$   &$\left(f^--1\right) f^+ x_2 x'_2$
    &$f^+ x_1 \left(\left(f^--1\right) x'_3+y\right)$  &$\left(f^--1\right) f^+ \bar x_1 x'_2$ \\

    $C_{a4}$    &$f^+ x_1 y$   &$f^+ \bar x_3 \left(\left(f^--1\right) x'_2+y\right)$
    &$f^+ \bar x_3 y$  &$f^+  \left(\left(f^--1\right) x'_2+y\right)$ \\

    $C_{a5}$    &$f^+ x_1 y$   &$f^+ \bar x_3 \left(\left(f^--1\right) x'_2+y\right)$
    &$f^+ x_1 \left(\left(f^--1\right) x'_2+y\right)$  &$f^+  \left(\left(f^--1\right) x'_2+y\right)$ \\

    $C_{a6}$    &$f^+ x_1 y$   &$\left(f^--1\right) f^+ x_2 x'_2$
    &$f^+ x_1$  &$\left(f^--1\right) f^+ \bar x_1 x'_2$ \\

     $C_{a7}$    &$f^+ x_1 y$   &$\left(f^--1\right) f^+ x_2 x'_2$
    &$r_c^2+\left(f^+ \left(x_1-1\right)+1\right) \left(\left(f^--1\right)
   x'_1+y\right)$  &$\left(f^--1\right) f^+ \bar x_1 x'_2$ \\

    $C_{b1}$    &$f^+ x_1 y$   &$\left(f^--1\right) f^+ x_2 x'_2$
    &$f^+ \left(x_1+x_3\right) y$  &$\left(f^--1\right) f^+ x_2 \bar x'_1$ \\

     $C_{b2}$    &$f^+ x_1 y$   &$\left(f^--1\right) f^+ x_2 x'_2$
    &$f^+ \bar x_3 \left(\left(f^--1\right)
   \bar x'_1+y\right)$  &$\left(f^--1\right) f^+ x_2 \bar x'_1$ \\

   $C_{b3}$    &$f^+ x_1 y$   &$\left(f^--1\right) f^+ x_2 x'_2$
    &$f^+ x_1 \left(\left(f^--1\right) x'_3+y\right)$  &$f^+ \bar x_3 \left(\left(f^--1\right)
   \bar x'_1+y\right)$ \\

    $C_{b4}$    &$f^+ x_1 y$   &$f^+ \bar x_3 \left(\left(f^--1\right) x'_2+y\right)$
    &$f^+ \bar x_3 y$  &$f^+ \bar x_3 \left(\left(f^--1\right)
   \bar x'_1+y\right)$ \\

   $C_{b5}$    &$f^+ x_1 y$   &$f^+ \bar x_3 \left(\left(f^--1\right) x'_2+y\right)$
    &$f^+ x_1 \left(\left(f^--1\right) x'_2+y\right)$  &$f^+ \bar x_3 \left(\left(f^--1\right)
   \bar x'_1+y\right)$ \\

    $C_{b6}$    &$f^+ x_1 y$   &$\left(f^--1\right) f^+ x_2 x'_2$
    &$f^+ x_1$  &$\left(f^--1\right) f^+ x_2 \bar x'_1$ \\

    $C_{b7}$    &$f^+ x_1 y$   &$\left(f^--1\right) f^+ x_2 x'_2$
    &$r_c^2+\left(f^+ \left(x_1-1\right)+1\right) \left(\left(f^--1\right)
   x'_1+y\right)$  &$\left(f^--1\right) f^+ x_2 \bar x'_1$ \\

   $C_{c1}$    &$f^+ x_1 y$   &$\left(f^--1\right) f^+ x_2 x'_2$
   &$f^+ \left(x_1+x_3\right) y$
   &\tabincell{l}{$r_c^2+f^+ \left(x_2-1\right) \left(x'_3-1\right)$\\$-f^- \left(f^+
   \left(x_2-1\right)+1\right) \left(x'_3-1\right)+x'_3-1$}\\

    $C_{c2}$    &$f^+ x_1 y$   &$\left(f^--1\right) f^+ x_2 x'_2$
   &$f^+ x_1 \left(\left(f^--1\right) x'_3+y\right)$
   &\tabincell{l}{$r_c^2+f^+ \left(x_2-1\right) \left(x'_3-1\right)$\\$-f^- \left(f^+
   \left(x_2-1\right)+1\right) \left(x'_3-1\right)+x'_3-1$}\\

   $C_{c3}$    &$f^+ x_1 y$   &$f^+ \bar x_3 \left(\left(f^--1\right) x'_2+y\right)$
    &$f^+ \bar x_3 y$   &\tabincell{l}{$r_c^2-\left(f^+ x_3-1\right) \left(x'_3+y-1\right)$\\$+f^- \left(f^+ x_3-1\right) \left(x'_3-1\right)$}\\

    $C_{c4}$    &$f^+ x_1 y$   &$f^+ \bar x_3 \left(\left(f^--1\right) x'_2+y\right)$
    &$f^+ x_1 \left(y-\left(f^--1\right) \left(x'_1+x'_3-1\right)\right)$
    &\tabincell{l}{$r_c^2-\left(f^+ x_3-1\right) \left(x'_3+y-1\right)$\\$+f^- \left(f^+ x_3-1\right) \left(x'_3-1\right)$}\\

   $C_{c5}$    &$f^+ x_1 y$   &$\left(f^--1\right) f^+ x_2 x'_2$
   &$f^+ x_1 $
   &\tabincell{l}{$r_c^2+f^+ \left(x_2-1\right) \left(x'_3-1\right)$\\$-f^- \left(f^+
   \left(x_2-1\right)+1\right) \left(x'_3-1\right)+x'_3-1$}\\

    $C_{c6}$    &$f^+ x_1 y$   &$\left(f^--1\right) f^+ x_2 x'_2$
   &\tabincell{l}{$r_c^2-\left(f^+ x_3-1\right) \left(x'_3+y-1\right)$\\$+f^- \left(f^+
   x_3-1\right) \left(x'_3-1\right)$}
   &\tabincell{l}{$r_c^2+f^+ \left(x_2-1\right) \left(x'_3-1\right)$\\$-f^- \left(f^+
   \left(x_2-1\right)+1\right) \left(x'_3-1\right)+x'_3-1$}\\

   $C_{c7}$    &$f^+ x_1 y$   &$\left(f^--1\right) f^+ x_2 x'_2$
   &$r_c^2+\left(f^+ \left(x_1-1\right)+1\right) \left(\left(f^--1\right) x'_1+y\right)$
   &\tabincell{l}{$r_c^2-\left(f^+ x_3-1\right) \left(x'_3+y-1\right)$\\$+f^- \left(f^+ x_3-1\right) \left(x'_3-1\right)$}\\

   $C_{d1}$    &$f^+ x_1 y$   &$\left(f^--1\right) f^+ x_2 x'_2$
    &$f^+ \left(x_1+x_3\right) y$  &$-f^- x'_2+f^+ x_2 \left(\left(f^--1\right) x'_2-y+1\right)+x'_2+y$ \\

    $C_{d2}$    &$f^+ x_1 y$   &$\left(f^--1\right) f^+ x_2 x'_2$
    &$f^+ x_1 \left(\left(f^--1\right) x'_3+y\right)$  &$-f^- x'_2+f^+ x_2 \left(\left(f^--1\right) x'_2-y+1\right)+x'_2+y$ \\

    $C_{d3}$    &$f^+ x_1 y$   &$f^+ \bar x_3 \left(\left(f^--1\right) x'_2+y\right)$
    &$f^+ \bar x_3 y$  &$-f^- x'_2-f^+ \left(x_3-1\right) \left(\left(f^--1\right)
   x'_2+1\right)+x'_2$ \\

   $C_{d4}$    &$f^+ x_1 y$   &$f^+ \bar x_3 \left(\left(f^--1\right) x'_2+y\right)$
    &$f^+ x_1 \left(\left(f^--1\right) x'_2+y\right)$  &$-f^- x'_2-f^+ \left(x_3-1\right) \left(\left(f^--1\right)
   x'_2+1\right)+x'_2$ \\

    $C_{d5}$    &$f^+ x_1 y$   &$\left(f^--1\right) f^+ x_2 x'_2$
    &$f^+ x_1$  &$-f^- x'_2-f^+ \left(x_3-1\right) \left(\left(f^--1\right)
   x'_2+1\right)+x'_2$ \\

   $C_{d6}$    &$f^+ x_1 y$
   &\tabincell{l}{$(1-f^-) x'_2+$\\$f^+ \bar x_3 \left(\left(f^--1\right)x'_2+1\right)$}
    &$f^+ x_1$  &$-f^- x'_2+f^+ x_2 \left(\left(f^--1\right) x'_2-y+1\right)+x'_2+y$ \\

     $C_{d7}$    &$f^+ x_1 y$   &$\left(f^--1\right) f^+ x_2 x'_2$
    &$r_c^2+\left(f^+ \left(x_1-1\right)+1\right) \left(\left(f^--1\right)
   x'_1+y\right)$  &$-f^- x'_2+f^+ x_2 \left(\left(f^--1\right) x'_2-y+1\right)+x'_2+y$ \\

 $C_{e1}$    &$\left(f^--1\right) f^+ x_2 x'_2$   &$f^+ \bar x_3 \left(\left(f^--1\right) x'_2+y\right)$
    &$f^+ x_2 \left(\left(f^--1\right) x'_2+y\right)$  &$f^+ \left(\left(f^--1\right) x'_2+y\right)$ \\

    $C_{e2}$    &$\left(f^--1\right) f^+ x_2 x'_2$   &$f^+ \bar x_3 \left(\left(f^--1\right) x'_2+y\right)$
    &$f^+ x_2 \left(\left(f^--1\right) x'_2+y\right)$  &$f^+ \bar x_3 \left(\left(f^--1\right)
   \bar x'_1+y\right)$ \\

   $C_{e3}$    &$\left(f^--1\right) f^+ x_2 x'_2$   &$f^+ \bar x_3 \left(\left(f^--1\right) x'_2+y\right)$
    &$f^+ x_2 \left(\left(f^--1\right) x'_2+y\right)$  &$-f^- x'_2+f^+ \bar x_3 \left(\left(f^--1\right)
   x'_2+1\right)+x'_2$ \\

    $C_{e4}$    &$\left(f^--1\right) f^+ x_2 x'_2$   &$f^+ \bar x_3 \left(\left(f^--1\right) x'_2+y\right)$
    &$f^+ x_2 \left(\left(f^--1\right) x'_2+y\right)$  &\tabincell{l}{$r_c^2-\left(f^+ x_3-1\right) \left(x'_3+y-1\right)$\\$+f^- \left(f^+ x_3-1\right) \left(x'_3-1\right)$}\\

    $C_{f1}$    &$\left(f^--1\right) f^+ x_2 x'_2$   &$f^+ \bar x_3 \left(\left(f^--1\right) x'_2+y\right)$
    &$\left(f^--1\right) f^+ \bar x_3 x'_2$  &$f^+ \left(\left(f^--1\right) x'_2+y\right)$ \\

    $C_{f2}$    &$\left(f^--1\right) f^+ x_2 x'_2$   &$f^+ \bar x_3 \left(\left(f^--1\right) x'_2+y\right)$
    &$\left(f^--1\right) f^+ \bar x_3 x'_2$  &$f^+ \left(-\left(x_3-1\right)\right) \left(\left(f^--1\right)
   \bar x'_1+y\right)$ \\

   $C_{f3}$    &$\left(f^--1\right) f^+ x_2 x'_2$   &$f^+ \bar x_3 \left(\left(f^--1\right) x'_2+y\right)$
    &$\left(f^--1\right) f^+ \bar x_3 x'_2$  &$-f^- x'_2-f^+ \left(x_3-1\right) \left(\left(f^--1\right)
   x'_2+1\right)+x'_2$ \\

   $C_{f4}$    &$\left(f^--1\right) f^+ x_2 x'_2$   &$f^+ \bar x_3 \left(\left(f^--1\right) x'_2+y\right)$
    &$\left(f^--1\right) f^+ \bar x_3 x'_2$  &\tabincell{l}{$r_c^2-\left(f^+ x_3-1\right) \left(x'_3+y-1\right)$\\$+f^- \left(f^+ x_3-1\right) \left(x'_3-1\right)$}\\

    $C_{g2}$  &$f^+ x_1 y$  &$\left(f^--1\right) f^+ x_2 x'_2$
    &$f^+ \bar x_3 \left(\left(f^--1\right) x'_2+y\right)$  &$f^+ \bar x_3 \left(\left(f^--1\right)
   \bar x'_1+y\right)$ \\

   $C_{g3}$  &$f^+ x_1 y$  &$\left(f^--1\right) f^+ x_2 x'_2$
    &$f^+ \bar x_3 \left(\left(f^--1\right) x'_2+y\right)$  &$-f^- x'_2+f^+ \bar x_3 \left(\left(f^--1\right)
   x'_2+1\right)+x'_2$ \\

   $C_{g4}$  &$f^+ x_1 y$  &$\left(f^--1\right) f^+ x_2 x'_2$
    &$f^+ \bar x_3 \left(\left(f^--1\right) x'_2+y\right)$   &\tabincell{l}{$r_c^2-\left(f^+ x_3-1\right) \left(x'_3+y-1\right)$\\$+f^- \left(f^+ x_3-1\right) \left(x'_3-1\right)$}\\

		\hline\hline
	\end{tabular}
\end{table*}

\begin{table*} [!htbh]	
	\scriptsize
	\centering
	\caption{Same as Table~\ref{tab:ttt} but  for the $W$-exchange diagrams.}
	\newcommand{\tabincell}[2]{\begin{tabular}{@{}#1@{}}#2\end{tabular}}
	\label{tab:ttte}
	\begin{tabular}[t]{lcccc}
		\hline\hline\
		$R_{ij}$ &  $\frac{t_A}{M^2}$   &$\frac{t_B}{M^2}$  &$\frac{t_C}{M^2}$ &$\frac{t_D}{M^2}$ \\ \hline
		
    $E_{a1}$    &$\left(f^--1\right) f^+ x_3 x'_3$   &$\left(f^--1\right) f^+ \bar x_1 \bar x'_1$
    &$\left(f^--1\right) f^+ \bar x_1 x'_3$  &$f^+ \left(-f^- x'_1+f^-+x'_1-1\right)$ \\	

    $E_{a2}$    &$\left(f^--1\right) f^+ x_3 x'_3$   &$\left(f^--1\right) f^+ \bar x_1 \bar x'_1$
    &$\left(f^--1\right) f^+ \bar x'_1 x_3$  &$\left(f^--1\right) f^+ \bar x'_1$ \\	

    $E_{a3}$    &$\left(f^--1\right) f^+ x_3 x'_3$   &$\left(f^--1\right) f^+ x_2 x'_2$
    &$\left(f^--1\right) f^+ \left(x_1+x_3\right) x'_3$  &$\left(f^--1\right) f^+ \bar x'_1$ \\	

    $E_{a4}$    &$\left(f^--1\right) f^+ x_3 x'_3$   &$\left(f^--1\right) f^+ x_2 x'_2$
    &$\left(f^--1\right) f^+ \bar x'_1$  &$\left(f^--1\right) f^+ \bar x_3 x'_2$ \\	

    $E_{a5}$    &$\left(f^--1\right) f^+ x_3 x'_3$   &$\left(f^--1\right) f^+ x_2 x'_2$
    &\tabincell{l}{$r_c^2+x'_2-1+f^+ \left(x_3-1\right) \left(x'_2-1\right)$\\$-f^- \left(f^+
   \left(x_3-1\right)+1\right) \left(x'_2-1\right)$}
    &$\left(f^--1\right) f^+ \bar x_3 x'_2$ \\	

    $E_{a6}$    &$\left(f^--1\right) f^+ x_3 x'_3$   &$\left(f^--1\right) f^+ x_2 x'_2$
   &$f^+ x_3 \left(\left(f^--1\right) x'_3+y\right)$
    &$\left(f^--1\right) f^+ \bar x_3 x'_2$ \\

    $E_{a7}$    &$\left(f^--1\right) f^+ x_3 x'_3$   &$\left(f^--1\right) f^+ x_2 x'_2$
     &\tabincell{l}{$x'_3+y-f^- x'_3$\\$+f^+ x_3 \left(\left(f^--1\right) x'_3-y+1\right)$}
    &$\left(f^--1\right) f^+ \bar x_3 x'_2$ \\		

     $E_{b1}$    &$\left(f^--1\right) f^+ x_3 x'_3$   &$\left(f^--1\right) f^+ \bar x_1 \bar x'_1$
    &$\left(f^--1\right) f^+ \bar x_1 x'_3$  &$r_c^2+f^-+f^+ x_1-f^- f^+ x_1-1$ \\	

     $E_{b2}$    &$\left(f^--1\right) f^+ x_3 x'_3$   &$\left(f^--1\right) f^+ \bar x_1 \bar x'_1$
    &$\left(f^--1\right) f^+ \bar x'_1 x_3$  &$r_c^2+f^-+f^+ x_1-f^- f^+ x_1-1$ \\	

    $E_{b3}$    &$\left(f^--1\right) f^+ x_3 x'_3$   &$\left(f^--1\right) f^+ x_2 x'_2$  &$\left(f^--1\right) f^+ \bar x'_1 x_3$
    &\tabincell{l}{$r_c^2-\left(f^+ \left(x_1+x_3\right)-1\right) \left(x'_3-1\right)$\\$+f^-
   \left(f^+ \left(x_1+x_3\right)-1\right) \left(x'_3-1\right)$} \\	

   $E_{b4}$    &$\left(f^--1\right) f^+ x_3 x'_3$   &$\left(f^--1\right) f^+ x_2 x'_2$  &$r_c^2+f^-+f^+ x_1-f^- f^+ x_1-1$
    &\tabincell{l}{$r_c^2+x'_3-1+f^+ \left(x_2-1\right) \left(x'_3-1\right)$\\$-f^- \left(f^+ \left(x_2-1\right)+1\right)
   \left(x'_3-1\right)$} \\	

    $E_{b5}$    &$\left(f^--1\right) f^+ x_3 x'_3$   &$\left(f^--1\right) f^+ x_2 x'_2$
    &\tabincell{l}{$r_c^2+x'_2-1+f^+ \left(x_3-1\right) \left(x'_2-1\right)$\\$-f^- \left(f^+ \left(x_3-1\right)+1\right)
   \left(x'_2-1\right)$}&$r_c^2+f^-+f^+ x_1-f^- f^+ x_1-1$ \\	

   $E_{b6}$    &$\left(f^--1\right) f^+ x_3 x'_3$   &$\left(f^--1\right) f^+ x_2 x'_2$  &$f^+ x_3 \left(\left(f^--1\right) x'_3+y\right)$
    &\tabincell{l}{$r_c^2+x'_3-1+f^+ \left(x_2-1\right) \left(x'_3-1\right)$\\$-f^- \left(f^+ \left(x_2-1\right)+1\right) \left(x'_3-1\right)$} \\	

    $E_{b7}$    &$\left(f^--1\right) f^+ x_3 x'_3$   &$\left(f^--1\right) f^+ x_2 x'_2$  &\tabincell{l}{$x'_3+y-f^- x'_3$\\$+f^+ x_3 \left(\left(f^--1\right) x'_3-y+1\right)$}
    &\tabincell{l}{$r_c^2+x'_3-1+f^+ \left(x_2-1\right) \left(x'_3-1\right)$\\$-f^- \left(f^+ \left(x_2-1\right)+1\right) \left(x'_3-1\right)$} \\	

    $E_{c1}$    &$\left(f^--1\right) f^+ x_3 x'_3$   &$\left(f^--1\right) f^+ \bar x_1 \bar x'_1$
    &$f^+ \left(-f^- x_1+f^-+x_1-1\right) x'_3$  &$f^+ \bar x_1 \left(\left(f^--1\right)
   \bar x'_1+y\right)$ \\	

   $E_{c2}$    &$\left(f^--1\right) f^+ x_3 x'_3$   &$\left(f^--1\right) f^+ \bar x_1 \bar x'_1$
    &$\left(f^--1\right) f^+ x_3 \bar x'_1$  &$f^+ \bar x_1 \left(\left(f^--1\right)
   \bar x'_1+y\right)$ \\	

   $E_{c3}$    &$\left(f^--1\right) f^+ x_3 x'_3$   &$\left(f^--1\right) f^+ x_2 x'_2$
    &$\left(f^--1\right) f^+ \left(x_1+x_3\right) x'_3$  &$f^+ x_2 \left(\left(f^--1\right) x'_2+y\right)$ \\

    $E_{c4}$    &$\left(f^--1\right) f^+ x_3 x'_3$   &$\left(f^--1\right) f^+ x_2 x'_2$
    &\tabincell{l}{$r_c^2+x'_2-1+f^+ \left(x_3-1\right) \left(x'_2-1\right)$\\$-f^- \left(f^+ \left(x_3-1\right)+1\right)
   \left(x'_2-1\right)$} &$f^+ x_2 \left(\left(f^--1\right) x'_2+y\right)$ \\

    $E_{c5}$    &$\left(f^--1\right) f^+ x_3 x'_3$   &$\left(f^--1\right) f^+ x_2 x'_2$
    &$f^+ x_3 \left(\left(f^--1\right) x'_3+y\right)$ &$f^+ \bar x_1 \left(\left(f^--1\right)
   \bar x'_1+y\right)$ \\

   $E_{c6}$    &$\left(f^--1\right) f^+ x_3 x'_3$   &$\left(f^--1\right) f^+ x_2 x'_2$
    &$f^+ \left(-\left(x_1-1\right)\right) \left(\left(f^--1\right) \bar x'_1+y\right)$ &$f^+ x_2 \left(\left(f^--1\right) x'_2+y\right)$ \\

    $E_{c7}$    &$\left(f^--1\right) f^+ x_3 x'_3$   &$\left(f^--1\right) f^+ x_2 x'_2$
    &$-f^- x'_3+f^+ x_3 \left(\left(f^--1\right) x'_3-y+1\right)+x'_3+y$ &$f^+ x_2 \left(\left(f^--1\right) x'_2+y\right)$ \\

     $E_{d1}$   &$\left(f^--1\right) f^+ x_3 x'_3$   &$\left(f^--1\right) f^+ \bar x_1 \bar x'_1$  &$f^+ \left(-f^- x_1+f^-+x_1-1\right) x'_3$
    &\tabincell{l}{$f^+ \bar x_1 \left(x'_1-y\right)-x'_1+y+1$\\$-f^- \left(f^+
   \bar x_1-1\right) \left(x'_1-1\right)$} \\	

   $E_{d2}$   &$\left(f^--1\right) f^+ x_3 x'_3$   &$\left(f^--1\right) f^+ \bar x_1 \bar x'_1$  &$\left(f^--1\right) f^+ x_3 \bar x'_1$
    &\tabincell{l}{$f^+ \bar x_1 \left(x'_1-y\right)-x'_1+y+1$\\$-f^- \left(f^+
   \bar x_1-1\right) \left(x'_1-1\right)$} \\	

    $E_{d3}$    &$\left(f^--1\right) f^+ x_3 x'_3$   &$\left(f^--1\right) f^+ x_2 x'_2$
    &$\left(f^--1\right) f^+ \left(x_1+x_3\right) x'_3$ &$-f^- x'_2+f^+ x_2 \left(\left(f^--1\right) x'_2-y+1\right)+x'_2+y$ \\

     $E_{d4}$    &$\left(f^--1\right) f^+ x_3 x'_3$   &$\left(f^--1\right) f^+ x_2 x'_2$
    &\tabincell{l}{$r_c^2+x'_2-1+f^+ \left(x_3-1\right) \left(x'_2-1\right)$\\$-f^- \left(f^+ \left(x_3-1\right)+1\right)
   \left(x'_2-1\right)$} &$-f^- x'_2+f^+ x_2 \left(\left(f^--1\right) x'_2-y+1\right)+x'_2+y$ \\

   $E_{d5}$    &$\left(f^--1\right) f^+ x_3 x'_3$   &$\left(f^--1\right) f^+ x_2 x'_2$
    &$f^+ x_3 \left(\left(f^--1\right) x'_3+y\right)$ &$-f^- x'_2+f^+ x_2 \left(\left(f^--1\right) x'_2-y+1\right)+x'_2+y$ \\

   $E_{d6}$    &$\left(f^--1\right) f^+ x_3 x'_3$   &$\left(f^--1\right) f^+ x_2 x'_2$
    &$y+(1-f^-) x'_3+f^+ x_3 \left(\left(f^--1\right) x'_3-y+1\right)$  &\tabincell{l}{$f^+ \bar x_1 \left(x'_1-y\right)-x'_1+y+1$\\$-f^- \left(f^+
   \bar x_1-1\right) \left(x'_1-1\right)$} \\

   $E_{d7}$    &$\left(f^--1\right) f^+ x_3 x'_3$   &$\left(f^--1\right) f^+ x_2 x'_2$
   &\tabincell{l}{$f^+ \bar x_1 \left(x'_1-y\right)-x'_1+y+1$\\$-f^- \left(f^+
   \bar x_1-1\right) \left(x'_1-1\right)$}
   &$-f^- x'_2+f^+ x_2 \left(\left(f^--1\right) x'_2-y+1\right)+x'_2+y$ \\

    $E_{e1}$    &$\left(f^--1\right) f^+ \bar x_1 \bar x'_1$   &$\left(f^--1\right) f^+ x_2 x'_2$
    &$\left(f^--1\right) f^+ \bar x'_1$
    &$f^+ x_2 \left(-f^- x'_1+f^-+x'_1-1\right)$ \\

    $E_{e2}$    &$\left(f^--1\right) f^+ \bar x_1 \bar x'_1$   &$\left(f^--1\right) f^+ x_2 x'_2$
    &$r_c^2+f^-+f^+ x_1-f^- f^+ x_1-1$
    &$f^+ x_2 \left(-f^- x'_1+f^-+x'_1-1\right)$ \\

    $E_{e3}$    &$\left(f^--1\right) f^+ \bar x_1 \bar x'_1$   &$\left(f^--1\right) f^+ x_2 x'_2$
     &\tabincell{l}{$f^+ \bar x_1 \left(x'_1-y\right)-x'_1+y+1$\\$-f^- \left(f^+
   \bar x_1-1\right) \left(x'_1-1\right)$}
    &$f^+ x_2 \left(-f^- x'_1+f^-+x'_1-1\right)$ \\

     $E_{e4}$    &$\left(f^--1\right) f^+ \bar x_1 \bar x'_1$   &$\left(f^--1\right) f^+ x_2 x'_2$
    &\tabincell{l}{$f^+ \bar x_1 \left(x'_1-y\right)-x'_1+y+1$\\$-f^- \left(f^+
   \bar x_1-1\right) \left(x'_1-1\right)$}
    &$f^+ x_2 \left(-f^- x'_1+f^-+x'_1-1\right)$ \\

    $E_{f1}$    &$\left(f^--1\right) f^+ \bar x_1 \bar x'_1$   &$\left(f^--1\right) f^+ x_2 x'_2$
    &$\left(f^--1\right) f^+ \bar x'_1$
    &$\left(f^--1\right) f^+ \bar x_1 x'_2$ \\	

    $E_{f2}$    &$\left(f^--1\right) f^+ \bar x_1 \bar x'_1$   &$\left(f^--1\right) f^+ x_2 x'_2$
    &$r_c^2+f^-+f^+ x_1-f^- f^+ x_1-1$
    &$\left(f^--1\right) f^+ \bar x_1 x'_2$ \\

    $E_{f3}$    &$\left(f^--1\right) f^+ \bar x_1 \bar x'_1$   &$\left(f^--1\right) f^+ x_2 x'_2$
    &$f^+ \left(-\left(x_1-1\right)\right) \left(\left(f^--1\right)
   \bar x'_1+y\right)$
    &$\left(f^--1\right) f^+ \bar x_1 x'_2$ \\

     $E_{f4}$    &$\left(f^--1\right) f^+ \bar x_1 \bar x'_1$   &$\left(f^--1\right) f^+ x_2 x'_2$
    &\tabincell{l}{$f^+ \bar x_1 \left(x'_1-y\right)-x'_1+y+1$\\$-f^- \left(f^+
   \bar x_1-1\right) \left(x'_1-1\right)$}
    &$\left(f^--1\right) f^+ \bar x_1 x'_2$ \\

		\hline\hline
	\end{tabular}
\end{table*}

In this Appendix, we present the explicit expressions for the quantities associated with individual diagrams contributing to the decay amplitude in Eq.~(\ref{eq:ab}) for the process $\bar B^0\rightarrow \Lambda_c^+ \bar p$. Table~\ref{tab:wilson} lists the combinations of Wilson coefficients $a_{R_{ij}}$. The virtualities of the internal propagators for the $W$-emission and $W$-exchange topologies are given in Tables~\ref{tab:ttt} and~\ref{tab:ttte}, respectively. 

The expressions for $H_{R_{ij}}(x_l,x'_l,y)$ in this channel are rather lengthy due to the inclusion of numerous higher-twist contributions. For brevity, we present only the results corresponding to the diagram in Fig.~\ref{fig:C}(d5), which provides the dominant contribution in this decay mode; the remaining terms can be obtained analogously.

For convenience, the amplitudes are organized according to the proton LCDAs defined in Eq.~(\ref{eq:LCDAsp}). For the $A$-term, we write
\begin{eqnarray}
H_{R_{ij}}^A(x_l,x'_l,y)=\sum_{\mathcal{F}_i} \mathcal{F}_i D_{\mathcal{F}_i}(x_l,x'_l,y),
\end{eqnarray}
with

\begin{widetext}

\begin{eqnarray}
D_{\mathcal{S}_1}(x_l,x'_l,y)&=&-\frac{4 M^4 \bar{r}}{f^--f^+} (-(\phi _3^a+\phi _3^s) x'_2 (f^+-1) (x_1 \bar{\phi }_B f^++\phi _B (1-2 x_1 f^+))
   (f^-)^2+\nonumber\\&&((\phi _B ((x_2+x_1 ((-r+\bar{r}-2) x'_2+5))
   (f^+)^2+((r-\bar{r}) x_2-r x'_2+\bar{r} x'_2+x'_2-x_1 (3 r-2 x'_2+\bar{r}\nonumber\\&& ((2 r+1) x'_2-3)+2)+2)
   f^++r+r \bar{r} x'_2-(\bar{r}+1) (x'_2+1))+\bar{\phi }_B (f^+ (r x'_2+x_1 (2 r+x'_2 (f^+-1)\nonumber\\&&-2
   f^++1)-x_2 f^+-1)+\bar{r} (x'_2 ((r x_1-1) f^++1)-2 x_1 f^+))) \phi _3^a+(\bar{\phi }_B
   (x_2+x_1 x'_2) (f^+)^2\nonumber\\&&+\bar{\phi }_B ((r-\bar{r}) x'_2+x_1 (-2 r-x'_2+\bar{r} (r
   x'_2+2)+1)+1) f^++\bar{r} \bar{\phi }_B x'_2-r \phi _B (x_2 f^++x'_2 f^+\nonumber\\&&+x_1 (x'_2 f^+-3) f^++\bar{r} x'_2 (2 x_1
   f^+-1)+1)+\phi _B (-x_1 (f^+)^2-x_2 (f^+)^2-2 x_1 f^+-x'_2 (f^+-1)\nonumber\\&& (2 x_1 f^+-1)+\bar{r}
   (-3 x_1 f^++x_2 f^++x'_2 (f^+-1) (x_1 f^++1)+1)-1)) \phi _3^s) f^-+(x_1 (\bar{\phi }_B\nonumber\\&&+\phi
   _B (x'_2 r-r+\bar{r}-\bar{r} x'_2-3)) \phi _3^a-x_1 (\bar{\phi }_B+\phi _B (-x'_2 r+r+\bar{r}
   (x'_2-1)-3)) \phi _3^s\nonumber\\&&-x_2 (\phi _3^a-\phi _3^s) (\phi _B-\bar{\phi }_B)) (f^+)^2+\bar{r}
   (-((r-1) \phi _B (x'_2+2)+\bar{\phi }_B (-r+x'_2+1)) \phi _3^a\nonumber\\&&-((r-1) \phi _B x'_2+\bar{\phi }_B
   (r+x'_2+1)) \phi _3^s)+((\bar{\phi }_B (-x'_2 r-r+\bar{r} (r x_2+x'_2+x_1 (-x'_2 r+2
   r+2)+1)+1)\nonumber\\&&+\phi _B (x'_2 r+r+\bar{r} (-r x_2+x_2-x'_2+x_1 (-5 r+(2 r+1) x'_2-4)-1)-1)) \phi
   _3^a\nonumber\\&&+(\phi _B (x'_2 r+r+\bar{r} ((r-1) x_2-x'_2+x_1 (r+(2 r+1) x'_2+2)-1)+1)\nonumber\\&&-\bar{\phi }_B (x'_2 r+r+\bar{r}
   (r x_2-x'_2+x_1 (r x'_2+2)-1)+1)) \phi _3^s) f^+),
   \end{eqnarray}
   \begin{eqnarray}
D_{\mathcal{S}_2}(x_l,x'_l,y)&=&\frac{8 M^4}{(f^--f^+) f^+} \bar{r}^2 (-\bar{r} \phi _B (\phi _3^a (-(f^--1) x'_2-r+1)+\phi
   _3^s (-(f^--1) x'_2+r+1))\nonumber\\&&+(f^+)^2 (x_1 (\phi _3^a
   \bar{\phi }_B (\bar{r} ((f^--1) (-x'_2)-1)+(f^--1) r
   x'_2-2 f^-+r+2)+\phi _3^a \nonumber\\&&\phi _B (2 \bar{r} ((f^--1) x'_2+1)-2
   (f^--1) r x'_2+3 f^--2 r-3)+\phi _B \phi _3^s (2 \bar{r} ((f^--1)
   x'_2+1)\nonumber\\&&-2 (f^--1) r x'_2-3 f^--2 r+3)+\phi _3^s \bar{\phi }_B (\bar{r}
   ((f^--1) (-x'_2)-1)+(f^--1) r x'_2+2
   f^-+r-2))\nonumber\\&&-(f^--1) x_2 \phi _B (\phi _3^a-\phi _3^s))+f^+ (x_1
   \bar{r} \phi _3^a \bar{\phi }_B ((f^--1) x'_2+2 r+1)\nonumber\\&&+\phi _3^a \phi _B (\bar{r}
   (-x_1 (2 (f^--1) x'_2+3 r+2)-f^- x'_2+r x_2+x'_2+1)+(f^--1) r
   x'_2-f^--r+1)\nonumber\\&&+x_1 \bar{r} \phi _3^s \bar{\phi }_B ((f^--1) x'_2-2 r+1)+\phi _B
   \phi _3^s (\bar{r} (x_1 (-2 (f^--1) x'_2+3 r-2)-f^- x'_2-r
   x_2+x'_2+1)\nonumber\\&&+(f^--1) r x'_2+f^--r-1))),
   \end{eqnarray}
   \begin{eqnarray}
D_{\mathcal{P}_1}(x_l,x'_l,y)&=&-\frac{4 M^4 }{f^--f^+}\bar{r} ((\phi _3^a+\phi _3^s) x'_2 (f^+-1) (x_1 \bar{\phi }_B
   f^++\phi _B (1-2 x_1 f^+)) (f^-)^2\nonumber\\&&+((\phi _B ((x_1
   ((-r+\bar{r}+2) x'_2-5)-x_2) (f^+)^2-((\bar{r}-r)
   x_2+r x'_2-\bar{r} x'_2+x'_2\nonumber\\&&+x_1 (3 r+2 x'_2+\bar{r} ((1-2 r) x'_2-3)-2)+2)
   f^++r-r \bar{r} x'_2-(\bar{r}-1) (x'_2+1))\nonumber\\&&+\bar{\phi }_B (f^+ (r
   x'_2+x_2 f^++x_1 (2 r-x'_2 (f^+-1)+2 f^+-1)+1)-\bar{r} (2 x_1 f^++x'_2
   (r x_1 f^++f^+-1)))) \phi _3^a\nonumber\\&&+(-\bar{\phi }_B (x_2+x_1 x'_2)
   (f^+)^2-\bar{\phi }_B ((\bar{r}-r) x'_2+x_1 (2 r-x'_2+\bar{r} (r
   x'_2-2)+1)+1) f^++\bar{r} \bar{\phi }_B x'_2-\nonumber\\&&r \phi _B (x_2 f^++x'_2 f^++x_1
   (x'_2 f^+-3) f^++\bar{r} x'_2 (1-2 x_1 f^+)+1)+\phi _B (x_1
   (f^+)^2+x_2 (f^+)^2+2 x_1 f^+\nonumber\\&&+x'_2 (f^+-1) (2 x_1
   f^+-1)+\bar{r} (-3 x_1 f^++x_2 f^++x'_2 (f^+-1) (x_1
   f^++1)+1)+1)) \phi _3^s) f^-\nonumber\\&&+(-x_1 (\bar{\phi }_B+\phi _B
   (-x'_2 r+r+\bar{r} (x'_2-1)-3)) \phi _3^a+x_1 (\bar{\phi }_B+\phi _B
   (x'_2 r-r+\bar{r}-\bar{r} x'_2-3)) \phi _3^s\nonumber\\&&+x_2 (\phi _3^a-\phi _3^s)
   (\phi _B-\bar{\phi }_B)) (f^+)^2-\bar{r} ((\bar{\phi }_B
   (r+x'_2+1)-(r+1) \phi _B (x'_2+2)) \phi _3^a+\nonumber\\&&(\bar{\phi }_B
   (-r+x'_2+1)-(r+1) \phi _B x'_2) \phi _3^s)+((\phi _B (x'_2
   r+r+\bar{r} ((r+1) x_2-x'_2\nonumber\\&&+x_1 (5 r+(1-2 r) x'_2-4)-1)+1)+\bar{\phi }_B
   (-x'_2 r-r+\bar{r} (-r x_2+x'_2+x_1 (x'_2 r-2 r+2)+1)-1)) \phi
   _3^a\nonumber\\&&+(\bar{\phi }_B (-x'_2 r-r+\bar{r} (r x_2+x'_2+x_1 (r
   x'_2-2)+1)+1)\nonumber\\&&-\phi _B (-x'_2 r-r+\bar{r} ((r+1) x_2+x'_2+x_1 (r+(2 r-1)
   x'_2-2)+1)+1)) \phi _3^s) f^+),
   \end{eqnarray}
   \begin{eqnarray}
D_{\mathcal{P}_2}(x_l,x'_l,y)&=&\frac{8 M^4}{(f^--f^+) f^+} \bar{r}^2 (\bar{r} \phi _B (\phi _3^s ((f^--1) x'_2+r-1)-\phi _3^a
   (-(f^--1) x'_2+r+1))\nonumber\\&&+(f^+)^2 (x_1 (\phi _3^a \phi _B
   (2 \bar{r} ((f^--1) x'_2+1)-2 (f^--1) r x'_2-3 f^--2 r+3)+\phi
   _3^a\nonumber\\&& \bar{\phi }_B (\bar{r} ((f^--1) (-x'_2)-1)+(f^--1) r
   x'_2+2 f^-+r-2)+\phi _3^s \bar{\phi }_B (\bar{r} ((f^--1)
   (-x'_2)-1)\nonumber\\&&+(f^--1) r x'_2-2 f^-+r+2)+\phi _B \phi _3^s (2 \bar{r}
   ((f^--1) x'_2+1)-2 (f^--1) r x'_2+3 f^--2
   r-3))\nonumber\\&&+(f^--1) x_2 \phi _B (\phi _3^a-\phi _3^s))+f^+ (x_1
   \bar{r} \phi _3^a \bar{\phi }_B ((f^--1) x'_2-2 r+1)+\phi _3^a \phi _B (\bar{r}
   (x_1 (-2 (f^--1) x'_2+3 r-2)\nonumber\\&&-f^- x'_2-r x_2+x'_2+1)+(f^--1) r
   x'_2+f^--r-1)+x_1 \bar{r} \phi _3^s \bar{\phi }_B ((f^--1) x'_2+2 r+1)\nonumber\\&&+\phi _B
   \phi _3^s (\bar{r} (-x_1 (2 (f^--1) x'_2+3 r+2)-f^- x'_2+r
   x_2+x'_2+1)+(f^--1) r x'_2-f^--r+1))),
   \end{eqnarray}
   \begin{eqnarray}
D_{\mathcal{V}_1}(x_l,x'_l,y)&=& \frac{4 M^4}{f^--f^+} ((\phi _2 (f^+-1) ((x_1+x_2-x'_2) f^+-1) (\phi _B
   (2 x_1 f^+-1)-x_1 \bar{\phi }_B f^+)+\phi _4 f^+ ((x_2-x'_2)\nonumber\\&& (\phi _B
   (r-\bar{r}+f^+)-\bar{\phi }_B f^+)+x_1 (\bar{\phi }_B (r-\bar{r}-f^+)+\phi
   _B (-r+\bar{r}+2 f^+)))) (f^-)^2+(x_1 \phi _2 (\phi _B
   ((\bar{r}-r) x_1\nonumber\\&&+(\bar{r}-r) x_2+r x'_2-\bar{r} x'_2+2 x'_2-2)-\bar{\phi
   }_B (x'_2-1)) (f^+)^3+(\bar{r} \phi _2 (r \bar{\phi }_B-(2 r+1) \phi
   _B) x_1^2+(\phi _4 (\bar{\phi }_B-2 \phi _B)\nonumber\\&&+\phi _2 (\phi _B (-(2 r+1)
   \bar{r} (x_2-x'_2)-2 x'_2+2)+\bar{\phi }_B (r+x'_2+\bar{r} (r x_2-r
   x'_2-1)-1))) x_1\nonumber\\&&-(x_2+1) \phi _4 (\phi _B-\bar{\phi }_B)-r
   \phi _2 (\phi _B-\bar{\phi }_B) (x_2-x'_2)+\bar{r} \phi _2 (\phi _B-\bar{\phi
   }_B) (x_2-x'_2)\nonumber\\&&-(\phi _2 \phi _B+2 \phi _4 (\bar{\phi }_B-\phi _B))
   x'_2) (f^+)^2+(\phi _2 (\bar{\phi }_B (\bar{r} (-(r-1)
   x_1+x_2-x'_2+1)-r)\nonumber\\&&+\phi _B (r+x'_2+\bar{r} (3 r x_1+(r-1) x_2-r
   x'_2+x'_2-1)))+\phi _4 (\bar{r} \bar{\phi }_B ((r+1) x_1+r
   (x_2-x'_2))\nonumber\\&&+\phi _B (r (x'_2-1)+\bar{r} (-(2 r+1) x_1-r x_2+x_2+r
   x'_2-2 x'_2+1)))) f^+-\bar{r} \phi _2 ((r-1) \phi _B+\bar{\phi
   }_B)) f^-\nonumber\\&&+f^+ ((\bar{r}-r) x_1 \phi _2 \phi _B (x'_2-1)
   (f^+)^2+((\phi _B-\bar{\phi }_B) (-\phi _4 (x'_2-1)-r \phi _2
   x'_2)\nonumber\\&&+\bar{r} \phi _2 (x_1 (r \bar{\phi }_B-(2 r+1) \phi _B)
   (x'_2-1)+(\phi _B-\bar{\phi }_B) x'_2)) f^++\bar{r} (\phi _2
   ((r-1) \phi _B\nonumber\\&&+\bar{\phi }_B) x'_2-\phi _4 ((r-1) \phi _B-r \bar{\phi }_B)
   (x'_2-1)))),
\end{eqnarray}
\begin{eqnarray}
D_{\mathcal{V}_2}(x_l,x'_l,y)&=&
-\frac{8 M^4}{(f^--f^+) f^+} \bar{r} (f^+ ((f^+)^2 x_1 \phi _2 (\bar{r}-r) (x'_2-1)
   (2 \phi _B-\bar{\phi }_B)-f^+ (\phi _2 \bar{r} (x_1 (x'_2-1) (2 \phi
   _B-\bar{\phi }_B)\nonumber\\&&+\phi _B x'_2)+\phi _B (-r \phi _2 x'_2-\phi _4
   (x'_2-1)))+\bar{r} \phi _B (r \phi _4 (x'_2-1)+\phi _2
   x'_2))+f^- ((f^+)^3 x_1 \phi _2 (\bar{r}-r)\nonumber\\&&
   (-x'_2+x_1+x_2) (2 \phi _B-\bar{\phi }_B)+(f^+)^2 (x_1 (\phi _2
   (\bar{\phi }_B (\bar{r} (-x'_2+x_2+1)-r)\nonumber\\&&+\phi _B (\bar{r} (2 x'_2-2
   x_2-3)+3 r))+\phi _4 (\bar{\phi }_B-\phi _B))+\phi _B (\phi _2
   (\bar{r}-r) x'_2+x_2 (\phi _2 (r-\bar{r})+\phi _4)+\phi _4 (1-2
   x'_2))\nonumber\\&&+x_1^2 \phi _2 \bar{r} (\bar{\phi }_B-2 \phi _B))+f^+ (\phi _2
   (\phi _B (\bar{r} (-x'_2+3 x_1+x_2+1)-r)-x_1 \bar{r} \bar{\phi }_B)+r \phi
   _4 \bar{r} (x_1 (\bar{\phi }_B-\phi _B)\nonumber\\&&+\phi _B (x_2-x'_2)))-\phi _2
   \bar{r} \phi _B)+(f^-)^2 (f^+)^2 \phi _4 (x_1 (\phi _B-\bar{\phi
   }_B)+\phi _B (x'_2-x_2))),
\end{eqnarray}
\begin{eqnarray}
D_{\mathcal{V}_3}(x_l,x'_l,y)&=&
 \frac{8 M^4}{f^--f^+} \bar{r} (\phi _4 (\phi _B (-f^+ (x'_2-2)+f^- (f^+ (2 r x_1+r
   x_2+2 x'_2-2)+r)+f^+ (f^-)^2 (-x'_2))\nonumber\\&&-f^+ \bar{\phi }_B (f^-
   (r x_1+r x_2-1)+1))+\phi _2 (\bar{\phi }_B (f^+ r (1-f^+ x_1
   (x'_2-1))+f^- ((f^+)^2 (x_1 (r x'_2+2)+x_2)\nonumber\\&&-f^+
   (2 x_1+x_2+1)+1))-\phi _B (f^+ r ((1-2 f^+ x_1) x'_2+3 f^+
   x_1+1)+f^- ((f^+)^2 (x_1 (2 r x'_2-x_2+2)\nonumber\\&&-x_1^2+x_2)-f^+
   (r x'_2+2 x_1+x_2+2)+(f^+)^3 x_1 (x_1+x_2)+2)))+\bar{r}
   (\phi _2 (\phi _B ((f^+)^2 x_1 (r x_1+r x_2+3)\nonumber\\&&+f^+ ((2 r-3) x_1+r
   x_2+1)+(f^--1) (f^+-1) (2 f^+ x_1-1) x'_2-2 r-1)\nonumber\\&&-\bar{\phi }_B
   (f^+ (x_1 (-(f^--1) x'_2+2 r-1)+r x_2+1)+(f^+)^2 x_1
   ((f^--1) x'_2+1)-r-1))\nonumber\\&&+\phi _4 (\phi _B ((f^--1) r
   x'_2+2 f^+ x_1+f^+ x_2-f^- (2 f^+ x_1+f^+ x_2+1)+2 r+1)\nonumber\\&&-\bar{\phi }_B
   (-(f^--1) f^+ x_1-(f^--1) f^+ x_2+r)))),
\end{eqnarray}
\begin{eqnarray}
D_{\mathcal{V}_4}(x_l,x'_l,y)&=&
 \frac{8 M^4 }{(f^--f^+)
   f^+}\bar{r}^2 ((f^+)^3 x_1 (x_1+x_2) \phi _2 (\bar{r}-r) \phi
   _B+(f^+)^2 (-x_2 (r \phi _2+\phi _4) (\phi _B-\bar{\phi }_B)\nonumber\\&&+x_1
   (r \phi _2 \bar{\phi }_B+\phi _4 (\bar{\phi }_B-2 \phi _B))-\phi _2 \bar{r}
   (x_1^2 ((2 r+1) \phi _B-r \bar{\phi }_B)+x_1 (x_2 (-r \bar{\phi }_B+2 r \phi
   _B+\phi _B)+\bar{\phi }_B)\nonumber\\&&+x_2 (\bar{\phi }_B-\phi _B)))+f^+ (\phi
   _2 (\phi _B (\bar{r} (3 r x_1+(r-1) x_2-1)+r)+\bar{\phi }_B (\bar{r}
   (-(r-1) x_1+x_2+1)-r))\nonumber\\&&+\phi _4 \bar{r} (x_2 (r \bar{\phi }_B-(r-1) \phi
   _B)+x_1 ((r+1) \bar{\phi }_B-(2 r+1) \phi _B)))+f^- (f^+ \phi _4
   (x_2 \nonumber\\&&(\phi _B (-\bar{r}+f^++r)-f^+ \bar{\phi }_B)+x_1 (\bar{\phi }_B
   (-\bar{r}-f^++r)+\phi _B (\bar{r}+2 f^+-r)))\nonumber\\&&+(f^+-1) \phi _2
   (f^+ (x_1+x_2)-1)(\phi _B (2 f^+ x_1-1)-f^+ x_1 \bar{\phi
   }_B))-\phi _2 \bar{r} (\bar{\phi }_B+(r-1) \phi _B)),
\end{eqnarray}
\begin{eqnarray}
D_{\mathcal{V}_5}(x_l,x'_l,y)&=&
\frac{8 M^4 }{(f^--f^+) f^+}\bar{r}^2 ((f^+)^3 x_1 (x_1+x_2) \phi _2 (r-\bar{r}) \phi
   _B+(f^+)^2 (x_1^2 \phi _2 \bar{r} (r \bar{\phi }_B+(1-2 r) \phi _B)\nonumber\\&&+x_1
   (\phi _2 (\phi _B (\bar{r} (-2 r x_2+x_2-4)+4 r)+\bar{\phi }_B
   (\bar{r} (r x_2+3)-3 r))+\phi _4 (2 \phi _B-\bar{\phi
   }_B))+x_2 (\phi _B-\bar{\phi }_B)\nonumber\\&& (\phi _2 (r-\bar{r})+\phi
   _4))+f^+ (\phi _4 (\phi _B (\bar{r} ((2 r-1) x_1+(r+1)
   x_2)+2)+\bar{r} (-(r-1) x_1-r x_2) \bar{\phi }_B)\nonumber\\&&+\phi _2 (\phi _B
   (\bar{r} ((3 r+4) x_1+(r+1) x_2+3)-3 r)-\bar{\phi }_B (\bar{r} ((r+3)
   x_1+x_2+1)-r)))+f^- ((f^+-1) \nonumber\\&&\phi _2 (f^+
   (x_1+x_2)-1) (\phi _B (2 f^+ x_1-1)-f^+ x_1 \bar{\phi }_B)-f^+ \phi
   _4 (\phi _B (x_2 (\bar{r}+f^+-r)\nonumber\\&&+x_1 (-\bar{r}+2 f^++r)+2)-f^+
   (x_1+x_2) \bar{\phi }_B+x_1 (\bar{r}-r) \bar{\phi }_B))\nonumber\\&&+\bar{r}
   (\phi _2 (\bar{\phi }_B-(r+3) \phi _B)+2 r \phi _4 \phi
   _B)),
   \end{eqnarray}
   \begin{eqnarray}
D_{\mathcal{V}_6}(x_l,x'_l,y)&=& -\frac{16 M^4 }{(f^--f^+) (f^+)^2}\bar{r}^3 ((f^+)^3 x_1 (x_1+x_2) \phi _2 (\bar{r}-r)
   (2 \phi _B-\bar{\phi }_B)+f^+ (\phi _2 (\phi _B ((3 x_1+x_2+1)
   \bar{r}-r)\nonumber\\&&-x_1 \bar{r} \bar{\phi }_B)+r \phi _4 \bar{r} (x_1 (\bar{\phi }_B-\phi
   _B)+x_2 \phi _B))+(f^+)^2 (x_1 ((f^--1) \phi _4\nonumber\\&&
   (\phi _B-\bar{\phi }_B)+\phi _2 (\phi _B (3 r-(2 x_2+3)
   \bar{r})+\bar{\phi }_B ((x_2+1) \bar{r}-r)))\nonumber\\&&+x_2 \phi _B (\phi
   _2 (r-\bar{r})-(f^--1) \phi _4)+x_1^2 \phi _2 \bar{r} (\bar{\phi }_B -2
   \phi _B))-\phi _2 \bar{r} \phi _B),
   \end{eqnarray}
   \begin{eqnarray}
D_{\mathcal{A}_1}(x_l,x'_l,y)&=&\frac{4 M^4 }{f^--f^+}((\phi _2 (f^+-1) ((x_1+x_2-x'_2) f^+-1) (\phi _B
   (2 x_1 f^+-1)-x_1 \bar{\phi }_B f^+)\nonumber\\&&+\phi _4 f^+ (x_1 (\phi _B (r-\bar{r}+2
   f^+)-\bar{\phi }_B (r-\bar{r}+f^+))-(x_2-x'_2) (\phi _B
   (r-\bar{r}-f^+)+\bar{\phi }_B f^+))) (f^-)^2\nonumber\\&&-(x_1 \phi _2
   (\phi _B (\bar{r} (x_1+x_2-x'_2)-2 x'_2+2)-r \phi _B
   (x_1+x_2-x'_2)+\bar{\phi }_B (x'_2-1)) (f^+)^3\nonumber\\&&-(\bar{r} \phi _2
   ((1-2 r) \phi _B+r \bar{\phi }_B) x_1^2+(\phi _4 (\bar{\phi }_B-2 \phi _B)+\phi
   _2 (\phi _B (-(2 r-1) \bar{r} (x_2-x'_2)-2 x'_2+2)\nonumber\\&&+\bar{\phi }_B
   (-r+x'_2+\bar{r} (r x_2-r x'_2+1)-1))) x_1-(x_2+1) \phi _4
   (\phi _B-\bar{\phi }_B)+r \phi _2 (\phi _B-\bar{\phi }_B)
   (x_2-x'_2)\nonumber\\&&-\bar{r} \phi _2 (\phi _B-\bar{\phi }_B) (x_2-x'_2)-(\phi
   _2 \phi _B+2 \phi _4 (\bar{\phi }_B-\phi _B)) x'_2) (f^+)^2+(\phi _4
   (\bar{r} \bar{\phi }_B (r (x'_2-x_2)-(r-1) x_1)\nonumber\\&&+\phi _B (r
   (x'_2-1)+\bar{r} ((2 r-1) x_1+(r+1) x_2-r x'_2-2 x'_2+1)))+\phi _2
   (\bar{\phi }_B (\bar{r} ((r+1) x_1+x_2-x'_2+1)\nonumber\\&&-r)-\phi _B (-r+x'_2+\bar{r}
   (3 r x_1+(r+1) x_2-r x'_2-x'_2+1)))) f^++\bar{r} \phi _2 ((r+1) \phi
   _B-\bar{\phi }_B)) f^-\nonumber\\&&+f^+ ((r-\bar{r}) x_1 \phi _2 \phi _B (x'_2-1)
   (f^+)^2+(\bar{r} \phi _2 (x_1 ((1-2 r) \phi _B+r \bar{\phi }_B)
   (x'_2-1)\nonumber\\&&-(\phi _B-\bar{\phi }_B) x'_2)-(\phi _B-\bar{\phi }_B)
   (\phi _4 (x'_2-1)-r \phi _2 x'_2)) f^++\bar{r} (\phi _2 ((r+1) \phi
   _B-\bar{\phi }_B) x'_2\nonumber\\&&-\phi _4 ((r+1) \phi _B-r \bar{\phi }_B)
   (x'_2-1)))),
   \end{eqnarray}
   \begin{eqnarray}
D_{\mathcal{A}_2}(x_l,x'_l,y)&=&-\frac{8 M^4 }{(f^--f^+) f^+}\bar{r} (f^+ ((f^+)^2 x_1 \phi _2 (r-\bar{r}) (x'_2-1)
   (2 \phi _B-\bar{\phi }_B)+f^+ (\phi _2 \bar{r} (x_1 (x'_2-1) (2 \phi
   _B-\bar{\phi }_B)\nonumber\\&&+\phi _B x'_2)+\phi _B (\phi _4 (x'_2-1)-r \phi _2
   x'_2))+\bar{r} \phi _B (r \phi _4 (x'_2-1)-\phi _2 x'_2))+f^-
   ((f^+)^3 x_1 \phi _2 (r-\bar{r})\nonumber\\&& (-x'_2+x_1+x_2) (2 \phi
   _B-\bar{\phi }_B)+(f^+)^2 (x_1 (\phi _2 (\phi _B (\bar{r} (-2
   x'_2+2 x_2+3)-3 r)+\bar{\phi }_B (\bar{r} (x'_2-x_2-1)+r))\nonumber\\&&+\phi _4
   (\bar{\phi }_B-\phi _B))+\phi _B (\phi _2 (r-\bar{r}) x'_2+x_2 (\phi
   _2 (\bar{r}-r)+\phi _4)+\phi _4 (1-2 x'_2))+x_1^2 \phi _2 \bar{r} (2
   \phi _B-\bar{\phi }_B))\nonumber\\&&+f^+ (\phi _2 (\phi _B (r-\bar{r} (-x'_2+3
   x_1+x_2+1))+x_1 \bar{r} \bar{\phi }_B)+r \phi _4 \bar{r} (x_1 (\bar{\phi
   }_B-\phi _B)+\phi _B (x_2-x'_2)))\nonumber\\&&+\phi _2 \bar{r} \phi
   _B)+(f^-)^2 (f^+)^2 \phi _4 (x_1 (\phi _B-\bar{\phi }_B)+\phi
   _B (x'_2-x_2))),
   \end{eqnarray}
   \begin{eqnarray}
D_{\mathcal{A}_3}(x_l,x'_l,y)&=&\frac{8 M^4}{f^--f^+} \bar{r} (-\phi _4 (\phi _B (f^+ (x'_2-2)+f^- (f^+ (2 r x_1+r
   x_2-2 x'_2+2)+r)+f^+ (f^-)^2 x'_2)\nonumber\\&&-f^+ \bar{\phi }_B (f^- (r x_1+r
   x_2+1)-1))+\phi _2 (\phi _B (f^+ r ((1-2 f^+ x_1) x'_2+3 f^+
   x_1+1)+f^- ((f^+)^2 \nonumber\\&&(x_1 (2 r x'_2+x_2-2)+x_1^2-x_2)+f^+
   (-r x'_2+2 x_1+x_2+2)+(f^+)^3 (-x_1)
   (x_1+x_2)-2))\nonumber\\&&-\bar{\phi }_B (f^+ r (1-f^+ x_1
   (x'_2-1))+f^- ((f^+)^2 (x_1 (r x'_2-2)-x_2)+f^+
   (2 x_1+x_2+1)-1)))\nonumber\\&&+\bar{r} (\phi _2 (\bar{\phi }_B (-f^+
   (x_1 ((f^--1) x'_2+2 r+1)+r x_2-1)+(f^+)^2 x_1
   ((f^--1) x'_2+1)+r-1)\nonumber\\&&+\phi _B ((f^+)^2 x_1 (r x_1+r
   x_2-3)+f^+ ((2 r+3) x_1+r x_2-1)-(f^--1) (f^+-1) (2 f^+
   x_1-1) x'_2\nonumber\\&&-2 r+1))+\phi _4 (\phi _B ((f^--1) r x'_2-2 f^+ x_1-f^+
   x_2+f^- (2 f^+ x_1+f^+ x_2+1)+2 r-1)\nonumber\\&&-\bar{\phi }_B ((f^--1) f^+
   x_1+(f^--1) f^+ x_2+r)))),
   \end{eqnarray}
   \begin{eqnarray}
D_{\mathcal{A}_4}(x_l,x'_l,y)&=&\frac{8 M^4}{(f^--f^+)
   f^+} \bar{r}^2 ((f^+)^3 x_1 (x_1+x_2) \phi _2 (r-\bar{r}) \phi
   _B+(f^+)^2 (x_2 (r \phi _2-\phi _4) (\phi _B-\bar{\phi }_B)\nonumber\\&&+x_1
   (\phi _4 (\bar{\phi }_B-2 \phi _B)-r \phi _2 \bar{\phi }_B)+\phi _2 \bar{r}
   (x_1^2 (r \bar{\phi }_B+(1-2 r) \phi _B)+x_1 (x_2 (r \bar{\phi }_B-2 r \phi
   _B+\phi _B)+\bar{\phi }_B)\nonumber\\&&+x_2 (\bar{\phi }_B-\phi _B)))+f^+ \bar{r}
   (\phi _4 (x_1 (\phi _B-\bar{\phi }_B)-x_2 \phi _B)+\phi _2
   ((x_2+1) \phi _B-(x_1+x_2+1) \bar{\phi }_B))\nonumber\\&&+f^+ r (\phi _2
   (\phi _B ((3 x_1+x_2) \bar{r}-1)+\bar{\phi }_B (1-x_1
   \bar{r}))+\phi _4 \bar{r} (x_1 (\bar{\phi }_B-2 \phi _B)+x_2 (\bar{\phi
   }_B-\phi _B)))\nonumber\\&&+f^- (f^+ \phi _4 (x_2 (\phi _B
   (\bar{r}+f^+-r)-f^+ \bar{\phi }_B)+x_1 (\phi _B (-\bar{r}+2
   f^++r)-\bar{\phi }_B (-\bar{r}+f^++r)))\nonumber\\&&+(f^+-1) \phi _2 (f^+
   (x_1+x_2)-1) (\phi _B (2 f^+ x_1-1)-f^+ x_1 \bar{\phi
   }_B))+\phi _2 \bar{r} (\bar{\phi }_B-(r+1) \phi _B)),
   \end{eqnarray}
   \begin{eqnarray}
D_{\mathcal{A}_5}(x_l,x'_l,y)&=& \frac{8 M^4}{(f^--f^+) f^+} \bar{r}^2 ((f^+)^3 x_1 (x_1+x_2) \phi _2 (\bar{r}-r) \phi
   _B+(f^+)^2 (x_1^2 \phi _2 (-\bar{r}) ((2 r+1) \phi _B-r \bar{\phi
   }_B)\nonumber\\&&-x_1 (\phi _2 (\phi _B (\bar{r} (2 r x_2+x_2-4)+4 r)-\bar{\phi
   }_B (\bar{r} (r x_2-3)+3 r))+\phi _4 (\bar{\phi }_B-2 \phi
   _B))\nonumber\\&&-x_2 (\phi _B-\bar{\phi }_B) (\phi _2 (r-\bar{r})-\phi
   _4))+f^+ (\phi _4 (\phi _B (\bar{r} ((2 r+1) x_1+(r-1)
   x_2)+2)-\bar{r} ((r+1) x_1+r x_2) \nonumber\\&&\bar{\phi }_B)+\phi _2 (\phi _B
   (\bar{r} ((3 r-4) x_1+(r-1) x_2-3)+3 r)+\bar{\phi }_B (\bar{r} (-(r-3)
   x_1+x_2+1)-r)))\nonumber\\&&+f^- (f^+ \phi _4 (\bar{\phi }_B (x_1
   (\bar{r}+f^+-r)+f^+ x_2)-\phi _B (x_2 (-\bar{r}+f^++r)+x_1 (\bar{r}+2
   f^+-r)+2))\nonumber\\&&+(f^+-1) \phi _2 (f^+ (x_1+x_2)-1) (\phi
   _B (2 f^+ x_1-1)-f^+ x_1 \bar{\phi }_B))\nonumber\\&&+\bar{r} (2 r \phi _4 \phi _B-\phi _2
   (\bar{\phi }_B+(r-3) \phi _B))),
   \end{eqnarray}
   \begin{eqnarray}
D_{\mathcal{A}_6}(x_l,x'_l,y)&=&\frac{16 M^4 }{(f^--f^+)
   (f^+)^2}\bar{r}^3 ((f^+)^3 x_1 (x_1+x_2) \phi _2 (\bar{r}-r)
   (2 \phi _B-\bar{\phi }_B)+f^+ (r \phi _4 \bar{r} (x_1 (\phi _B-\bar{\phi
   }_B)-x_2 \phi _B)\nonumber\\&&+\phi _2 (\phi _B ((3 x_1+x_2+1) \bar{r}-r)-x_1
   \bar{r} \bar{\phi }_B))+(f^+)^2 (x_1 (\phi _2 (\phi _B (3
   r-(2 x_2+3) \bar{r})+\bar{\phi }_B\nonumber\\&& ((x_2+1)
   \bar{r}-r))-(f^--1) \phi _4 (\phi _B-\bar{\phi }_B))+x_2 \phi _B
   (\phi _2 (r-\bar{r})+(f^--1) \phi _4)\nonumber\\&&+x_1^2 \phi _2 \bar{r}
   (\bar{\phi }_B-2 \phi _B))-\phi _2 \bar{r} \phi _B),
   \end{eqnarray}
   \begin{eqnarray}
D_{\mathcal{T}_1}(x_l,x'_l,y)&=&-\frac{4 M^4}{f^--f^+} ((x_1 \phi _B (2 x_1 (\phi _3^a-\phi _3^s)+2 x_2 (\phi _3^a-\phi
   _3^s)-(\phi _3^a+\phi _3^s) x'_2) (f^+)^3+(2 (r-\bar{r})
   (\phi _3^a-\phi _3^s)\nonumber\\&& (2 \phi _B-\bar{\phi }_B) x_1^2+((\bar{\phi }_B
   (-2 (r-\bar{r}) x_2+(r-\bar{r}) x'_2-2)+\phi _B (4
   (r-\bar{r}) x_2+(-2 r+2 \bar{r}+1) x'_2+1)) \phi _3^a\nonumber\\&&+(\phi _B
   (-4 (r-\bar{r}) x_2+(-2 r+2 \bar{r}+1) x'_2-1)-\bar{\phi }_B (-2
   (r-\bar{r}) x_2+(\bar{r}-r) x'_2-2)) \phi _3^s) x_1\nonumber\\&&+x_2 (\phi
   _3^a-\phi _3^s) (3 \phi _B-2 \bar{\phi }_B)-(\phi _3^a+\phi _3^s) (3 \phi
   _B-\bar{\phi }_B) x'_2) (f^+)^2+((\bar{\phi }_B (x_1 (2
   r+\bar{r} (x'_2-2))-x'_2+2)\nonumber\\&&+\phi _B (-2 (r-\bar{r}) x_2-2 x_1
   (3 r+\bar{r} (x'_2-3))+r x'_2-\bar{r} x'_2+3 x'_2-3)) \phi _3^a+(-2
   r x_1 \bar{\phi }_B+(\bar{r} x_1-1)\nonumber\\&& (x'_2+2) \bar{\phi }_B+r \phi _B (6 x_1+2
   x_2+x'_2)-\phi _B (\bar{r} (2 x_2+x'_2+2 x_1 (x'_2+3))-3
   (x'_2+1))) \phi _3^s) f^+\nonumber\\&&+\phi _B ((2 r+\bar{r}
   (x'_2-2)) \phi _3^a+(\bar{r} (x'_2+2)-2 r) \phi _3^s))
   (f^-)^2-(x_1 \phi _B (\phi _3^a (2 x_1+2 x_2-x'_2+1)\nonumber\\&&-\phi _3^s (2
   x_1+2 x_2+x'_2-1)) (f^+)^3-(\bar{\phi }_B (-2 \bar{r} x_1^2+(-x'_2
   r+2 r+\bar{r} (-2 x_2+x'_2-2)+2) x_1\nonumber\\&&+(r-\bar{r}+2) x_2-2 r x'_2+2 \bar{r}
   x'_2-x'_2+1) \phi _3^a+\phi _B (-2 (r-2) \bar{r} x_1^2+(2 x'_2 r-3 r-x'_2\nonumber\\&&+\bar{r} (-2
   (r-2) x_2+(r-2) x'_2+3)) x_1+(-r+\bar{r}-3) x_2+2 r x'_2-2 \bar{r} x'_2+3
   x'_2-3) \phi _3^a\nonumber\\&&+\bar{\phi }_B (2 \bar{r} x_1^2+(-r x'_2+\bar{r} (2
   x_2+x'_2)-2) x_1+(-r+\bar{r}-2) x_2-2 r x'_2+2 \bar{r} x'_2-x'_2+1) \phi
   _3^s\nonumber\\&&-\phi _B (-2 (r-2) \bar{r} x_1^2+(r+(1-2 r) x'_2-\bar{r} (2 (r-2) x_2+(r-2)
   x'_2+1)-2) x_1-r x_2\nonumber\\&&+(\bar{r}-3) x_2-2 r x'_2+2 \bar{r} x'_2-3 x'_2+3) \phi
   _3^s) (f^+)^2+((\phi _B ((r+3) (x'_2-2)+\bar{r} ((3 r+2)
   x_2\nonumber\\&&+x_1 (r-2 x'_2+8)-3 r x'_2-3 x'_2+2))+\bar{\phi }_B (r-x'_2+\bar{r} (-2
   r x_2+r x'_2+2 x'_2\nonumber\\&&+x_1 (-2 r+x'_2-3)-1)+3)) \phi _3^a+(-(r+1) \bar{\phi
   }_B+((r+3) \phi _B-\bar{\phi }_B) x'_2+\bar{r} (r x'_2 \bar{\phi }_B+2 x'_2 \bar{\phi
   }_B\nonumber\\&&+\bar{\phi }_B+x_2 (2 r \bar{\phi }_B-(3 r+2) \phi _B)-3 r \phi _B x'_2-3 \phi _B x'_2+x_1
   (\bar{\phi }_B (2 r+x'_2+1)-\phi _B (r+2 x'_2+4)))) \phi
   _3^s)\nonumber\\&& f^++\bar{r} ((2 r \bar{\phi }_B+\phi _B (-3 r+x'_2-3)) \phi
   _3^a+(\phi _B (3 r+x'_2+1)-2 r \bar{\phi }_B) \phi _3^s)) f^-\nonumber\\&&+f^+
   (\bar{r} (\bar{\phi }_B (-r+(r+2) x'_2-3)+\phi _B (3 (r+1)-(3 r+2)
   x'_2)) \phi _3^a+\bar{r} (\bar{\phi }_B (-r+(r+2) x'_2-1)\nonumber\\&&+\phi _B (3 r-(3
   r+2) x'_2+1)) \phi _3^s-r (\phi _3^a+\phi _3^s) ((\bar{r} x_1+2)
   \phi _B-2 \bar{\phi }_B) (x'_2-1) f^+\nonumber\\&&-\bar{r} (\phi _B-\bar{\phi }_B)
   (\phi _3^a (x_1+x_2-2 x'_2+2)-\phi _3^s (x_1+x_2+2 x'_2-2))
   f^+)),
   \end{eqnarray}
   \begin{eqnarray}
D_{\mathcal{T}_2}(x_l,x'_l,y)&=&\frac{4 M^4 }{f^--f^+}\bar{r} (-(\phi _3^a+\phi _3^s) x'_2 (f^+-1) (x_1 \bar{\phi }_B
   f^++\phi _B (1-2 x_1 f^+)) (f^-)^2\nonumber\\&&+((\bar{\phi }_B (x_2+x_1
   x'_2) (f^+)^2-\bar{\phi }_B ((\bar{r}-r) x'_2+x_1 ((1-r
   \bar{r}) x'_2-1)+1) f^++\bar{r} \bar{\phi }_B x'_2\nonumber\\&&-r \phi _B (x_2 f^++x'_2 f^++x_1
   (x'_2 f^++1) f^++\bar{r} x'_2 (2 x_1 f^+-1)-1)+\phi _B (x_1
   (f^+)^2-x_2 (f^+)^2\nonumber\\&&-2 x_1 f^+-x'_2 (f^+-1) (2 x_1
   f^+-1)+\bar{r} (x_1 f^++x_2 f^++x'_2 (f^+-1) (x_1
   f^++1)-1)+1)) \phi _3^a\nonumber\\&&+(\phi _B ((x_2+x_1
   ((-r+\bar{r}-2) x'_2+3)) (f^+)^2+((r-\bar{r}) x_2-r
   x'_2+\bar{r} x'_2+x'_2-x_1 (-r+\bar{r}\nonumber\\&&+(2 r+1) \bar{r} x'_2-2 x'_2+2)-2)
   f^+-(\bar{r}+1) (x'_2-1)+r (\bar{r} x'_2-1))+\bar{\phi }_B
   (\bar{r} x'_2 ((r x_1-1) f^++1)\nonumber\\&&+f^+ (r x'_2+x_1 (x'_2
   (f^+-1)-2 f^++1)-x_2 f^++1))) \phi _3^s) f^-+(x_2 (\phi
   _3^a-\phi _3^s) (\phi _B-\bar{\phi }_B)\nonumber\\&&+x_1 ((\phi _B (x'_2
   r-r+\bar{r}-\bar{r} x'_2+1)-\bar{\phi }_B) \phi _3^a+(\bar{\phi }_B+\phi _B (x'_2
   r-r+\bar{r}-\bar{r} x'_2-1)) \phi _3^s)) (f^+)^2\nonumber\\&&+\bar{r} ((r+1)
   \bar{\phi }_B \phi _3^a+(r-1) (2 \phi _B-\bar{\phi }_B) \phi _3^s-(\phi _3^a+\phi
   _3^s) ((r-1) \phi _B+\bar{\phi }_B) x'_2)\nonumber\\&&+(-\phi _B (-x'_2 r+r+\bar{r}
   (-r x_2+x_2+x'_2+x_1 (r-(2 r+1) x'_2+2)-1)+1) \phi _3^a\nonumber\\&&+\bar{\phi }_B
   (\bar{r} (x'_2-1)-r (x'_2+\bar{r} (x_2+x_1 x'_2)-1)+1) \phi
   _3^a+(\bar{\phi }_B (-x'_2 r+r+\bar{r} (r x_2-r x_1
   (x'_2-2)\nonumber\\&&+x'_2-1)-1)+\phi _B (x'_2 r-r+\bar{r} (-r x_2+x_2-x'_2+x_1
   ((2 r+1) x'_2-3 r)+1)+1)) \phi _3^s) f^+),
   \end{eqnarray}
   \begin{eqnarray}
D_{\mathcal{T}_3}(x_l,x'_l,y)&=&
   \frac{16 M^4}{f^--f^+} \bar{r} ((f^+)^2 x_1 \phi _B (-\bar{r} \phi _3^a (-x'_2+2 x_1+2
   x_2+1)+\bar{r} \phi _3^s (x'_2+2 x_1+2 x_2-1)-r (x'_2-1) (\phi _3^a+\phi
   _3^s))\nonumber\\&&+f^+ (-r (x'_2-1) (3 \phi _B-\bar{\phi }_B) (\phi
   _3^a+\phi _3^s)-\bar{r} (\phi _3^a ((x'_2-2 x_1-2 x_2-1) \bar{\phi }_B+\phi _B
   ((x_1-3) x'_2+3 x_2+3))\nonumber\\&&+\phi _3^s ((x'_2+2 x_1+2 x_2-1)
   \bar{\phi }_B+\phi _B (x_1 (x'_2-2)-3
   (x'_2+x_2-1)))))+f^- (\bar{r} (\phi _3^a (\bar{\phi }_B
   ((f^+-1) x'_2\nonumber\\&&-2 (f^+ x_1+f^+ x_2-1))+\phi _B ((2 f^+
   x_1+3) (f^+ x_1+f^+ x_2-1)-(f^+-1) (f^+ x_1+3)
   x'_2))\nonumber\\&&-\phi _3^s (\bar{\phi }_B (-f^+ (x'_2+2 x_1+2
   x_2)+x'_2+2)+\phi _B ((f^+-1) (f^+ x_1+3) x'_2+(2 f^+
   x_1+3) \nonumber\\&&(f^+ x_1+f^+ x_2-1))))-r (\phi _3^a (\bar{\phi }_B
   (f^+ (x'_2-2 x_1-2 x_2)+2)+\phi _B ((f^+)^2 x_1 (-x'_2+2 x_1+2
   x_2)\nonumber\\&&+f^+ (-3 x'_2+x_1+3 x_2)-3))+\phi _3^s \bar{\phi }_B (f^+
   (x'_2+2 x_1+2 x_2)-2)-\phi _B \phi _3^s ((f^+)^2 x_1 (x'_2+2 x_1+2
   x_2)\nonumber\\&&+f^+ (3 (x'_2+x_2)+x_1)-3)))-\bar{r} (\phi _3^a
   (3 \phi _B (x'_2-2)-(x'_2-3) \bar{\phi }_B)+\phi _3^s (3 \phi _B
   x'_2-(x'_2+1) \bar{\phi }_B))),
   \end{eqnarray}
   \begin{eqnarray}
D_{\mathcal{T}_4}(x_l,x'_l,y)&=& -\frac{4 M^4 }{(f^--f^+) f^+}\bar{r} (f^+ (\phi _B (2 x_1 f^+-1)-x_1 \bar{\phi }_B f^+)
   ((x'_2+(4 x_1+4 x_2-x'_2) f^+-4) \phi _3^a\nonumber\\&&+(x'_2-(4 x_1+4
   x_2+x'_2) f^++4) \phi _3^s) (f^-)^2+((-4 (\phi _3^a-\phi
   _3^s) (2 \phi _B-\bar{\phi }_B) x_1^2+(\bar{\phi }_B (4 x_2-x'_2+2) \phi
   _3^a\nonumber\\&&+\phi _B (-8 x_2+(-r+\bar{r}+2) x'_2-3) \phi _3^a-\bar{\phi }_B (4
   x_2+x'_2) \phi _3^s+\phi _B (8 x_2+(-r+\bar{r}+2) x'_2-1) \phi _3^s)
   x_1\nonumber\\&&-x_2 (\phi _3^a-\phi _3^s) (\phi _B-\bar{\phi }_B))
   (f^+)^3+((\phi _B (-8 r \bar{r} x_1^2-(-r+2 x'_2+\bar{r} (8 r x_2+(1-2
   r) x'_2+1)-14) x_1\nonumber\\&&+(r-\bar{r}+4) x_2-5 r x'_2+5 \bar{r} x'_2-x'_2+2)-\bar{\phi
   }_B (-4 r \bar{r} x_1^2+(-x'_2+r \bar{r} (x'_2-4 x_2)+5)
   x_1\nonumber\\&&+(\bar{r}-r) x'_2+1)) \phi _3^a+(4 r \bar{r} (2 \phi _B-\bar{\phi
   }_B) x_1^2+(\bar{\phi }_B (x'_2-r \bar{r} (4 x_2+x'_2)+3)\nonumber\\&&+\phi _B
   (-r-2 x'_2+\bar{r} (8 r x_2+(2 r-1) x'_2+1)-10)) x_1+(-r+\bar{r}-4)
   x_2 \phi _B+\bar{\phi }_B-5 r \phi _B x'_2\nonumber\\&&+5 \bar{r} \phi _B x'_2-\phi _B x'_2+r \bar{\phi }_B x'_2-\bar{r}
   \bar{\phi }_B x'_2) \phi _3^s) (f^+)^2+((\bar{r} \bar{\phi }_B
   (x'_2-4 r x_1)\nonumber\\&&+\phi _B (-r+x'_2+\bar{r} (12 r x_1+4 r x_2-r x'_2-5
   x'_2+1)-5)) \phi _3^a+(\bar{r} \bar{\phi }_B (4 r x_1+x'_2)\nonumber\\&&+\phi _B
   (r+x'_2-\bar{r} (12 r x_1+4 r x_2+r x'_2+5 x'_2+1)+3)) \phi _3^s) f^+-4 r
   \bar{r} (\phi _3^a-\phi _3^s) \phi _B) f^-\nonumber\\&&+f^+ (\bar{r} (\bar{\phi }_B
   (r-x'_2+1)+\phi _B ((r+5) x'_2-2 (r+3))) \phi _3^a+(\bar{\phi }_B
   (-x'_2 r+r\nonumber\\&&+\bar{r} (-r x_2+r x_1 (x'_2-2)+x'_2-1)+1)+\phi _B (5 x'_2
   r-5 r+\bar{r} ((r+1) x_2-5 x'_2\nonumber\\&&+x_1 (3 r+(1-2 r) x'_2)+5)-1)) f^+ \phi
   _3^a+\bar{r} (\phi _B ((r+5) x'_2-4)-\bar{\phi }_B (r+x'_2-1)) \phi
   _3^s\nonumber\\&&+(\bar{\phi }_B (-x'_2 r+r+\bar{r} (r x_2+(r x_1+1)
   x'_2-1)-1)+\phi _B (5 x'_2 r-5 r-\bar{r} (r x_2+x_2+5 x'_2\nonumber\\&&+x_1 (-r+(2 r-1)
   x'_2+2)-5)+1)) f^+ \phi _3^s+x_2 (\phi _3^a-\phi _3^s) (\phi
   _B-\bar{\phi }_B) (f^+)^2\nonumber\\&&+x_1 ((\phi _B (x'_2 r-r+\bar{r}-\bar{r}
   x'_2+1)-\bar{\phi }_B) \phi _3^a+(\bar{\phi }_B+\phi _B (x'_2 r-r+\bar{r}-\bar{r}
   x'_2-1)) \phi _3^s) (f^+)^2)),
   \end{eqnarray}
   \begin{eqnarray}
D_{\mathcal{T}_5}(x_l,x'_l,y)&=& \frac{8 M^4}{(f^--f^+) f^+} \bar{r}^2 (-2 x_1 (x_1+x_2) (\phi _3^a-\phi _3^s) \phi _B
   (f^--1) (f^+)^3+(2 (\phi _3^a-\phi _3^s) (\bar{r} ((r-2)
   \phi _B+\bar{\phi }_B)\nonumber\\&&+(\bar{r}-r) (2 \phi _B-\bar{\phi }_B) f^-)
   x_1^2+(\bar{\phi }_B (2 x_2 f^- r-r+\bar{r} (-2 x_2 (f^--1)\nonumber\\&&+x'_2
   (f^--1)+1)+2 f^-+x'_2 (r-r f^-)-2) \phi _3^a+\phi _B (2 x'_2
   (f^--1) r-4 x_2 f^- r+2 r-f^-\nonumber\\&&+2 \bar{r} (-x'_2 (f^--1)+x_2 (r+2
   f^--2)-1)+1) \phi _3^a+\bar{\phi }_B (r (x'_2-1)-\bar{r} (2
   x_2+x'_2-1)\nonumber\\&&+(-2 (r-\bar{r}) x_2+(\bar{r}-r) x'_2-2) f^-+2)
   \phi _3^s+\phi _B (2 x'_2 (f^--1) r+4 x_2 f^- r+2 r+f^-\nonumber\\&&-2 \bar{r} (x'_2
   (f^--1)+x_2 (r+2 f^--2)+1)-1) \phi _3^s) x_1-x_2 (\phi
   _3^a-\phi _3^s) (3 \phi _B-2 \bar{\phi }_B) (f^--1))
   (f^+)^2\nonumber\\&&+(\bar{\phi }_B (-\bar{r} (2 r x_2+x_1 (2 r+x'_2 (f^--1)-2
   f^-+3))-2 (r x_1 f^-+f^--1)) \phi _3^a\nonumber\\&&+\phi _B (6 x_1 f^- r+2 x_2 f^-
   r-r+3 f^-+x'_2 (r-r f^-)+\bar{r} (f^- x'_2-x'_2+x_1 (r+2 x'_2 (f^--1)-6
   f^-+8)\nonumber\\&&+x_2 (3 r-2 f^-+2)+1)-3) \phi _3^a+\bar{\phi }_B (\bar{r} (2 r
   x_2+x_1 (2 r-x'_2 (f^--1)-2 f^-+1))\nonumber\\&&+2 (r x_1 f^-+f^--1)) \phi
   _3^s+\phi _B (-6 x_1 f^- r-2 x_2 f^- r-r-3 f^-+x'_2 (r-r f^-)\nonumber\\&&-\bar{r} (-f^-
   x'_2+x'_2+x_1 (r-2 x'_2 (f^--1)-6 f^-+4)+x_2 (3 r-2
   f^-+2)-1)+3) \phi _3^s) f^+\nonumber\\&&+\bar{r} ((2 r \bar{\phi }_B+\phi _B (-3
   r+x'_2-3)) \phi _3^a+(\phi _B (3 r+x'_2+1)-2 r \bar{\phi }_B) \phi
   _3^s)\nonumber\\&&+\phi _B (\phi _3^s (2 r-\bar{r} (x'_2+2))-\phi _3^a (2
   r+\bar{r} (x'_2-2))) f^-),
   \end{eqnarray}
   \begin{eqnarray}
D_{\mathcal{T}_6}(x_l,x'_l,y)&=& \frac{8 M^4}{(f^--f^+) f^+} \bar{r}^2 (\bar{r} \phi _B (\phi _3^a ((f^--1) x'_2+r+1)+\phi _3^s
   ((f^--1) x'_2-r+1))\nonumber\\&&+(f^+)^2 (x_1 (-(\bar{r}-r)
   \phi _3^a \bar{\phi }_B ((f^--1) x'_2+1)+\phi _3^a \phi _B (2 \bar{r}
   ((f^--1) x'_2+1)\nonumber\\&&-2 (f^--1) r x'_2+f^--2 r-1)-(\bar{r}-r)
   \phi _3^s \bar{\phi }_B ((f^--1) x'_2+1)+\phi _B \phi _3^s (2 \bar{r}
   ((f^--1) x'_2+1)\nonumber\\&&-2 (f^--1) r x'_2-f^--2
   r+1))+(f^--1) x_2 \phi _B (\phi _3^a-\phi _3^s))-f^+ (x_1
   (-\bar{r}) \phi _3^a \bar{\phi }_B ((f^--1) x'_2+1)\nonumber\\&&+\phi _3^a \phi _B
   (\bar{r} (x_1 (2 (f^--1) x'_2+r+2)+f^- x'_2+r x_2-x'_2+1)+(r-f^-
   r) x'_2+f^--r-1)\nonumber\\&&-x_1 \bar{r} \phi _3^s \bar{\phi }_B ((f^--1)
   x'_2+1)+\phi _B \phi _3^s (f^- (x'_2 (2 x_1 \bar{r}+\bar{r}-r)-1)\nonumber\\&&-r
   ((x_1+x_2) \bar{r}-x'_2+1)-(2 x_1+1) \bar{r}
   (x'_2-1)+1))),
   \end{eqnarray}
    \begin{eqnarray}
D_{\mathcal{T}_7}(x_l,x'_l,y)&=& -\frac{32 M^4}{(f^--f^+) f^+} \bar{r}^2 (f^+ (x_1+x_2)-1) ((f^--1) f^+-r \bar{r})\nonumber\\&&
   (\phi _3^a-\phi _3^s) (\phi _B (2 f^+ x_1+3)-2 \bar{\phi}_B),
   \end{eqnarray}
   \begin{eqnarray}
D_{\mathcal{T}_8}(x_l,x'_l,y)&=&-\frac{32 M^4}{(f^--f^+)
   (f^+)^2} \bar{r}^3 (f^+ x_1+f^+ x_2-1) (r \bar{r}+f^+-f^- f^+) \nonumber\\&&(\phi _3^a-\phi
   _3^s) (f^+ x_1 \bar{\phi }_B+\phi _B (1-2 f^+ x_1)).
   \end{eqnarray}
The corresponding formulas for the $B$ one can be obtained by the replacement $H_{R_{ij}}^B=H_{R_{ij}}^A|_{r\rightarrow -r}$.
\end{widetext}

\end{appendix}
\bibliographystyle{bibstyle}
\bibliography{biblio}

\end{document}